\documentclass[12pt]{article}
\usepackage{graphicx}
\usepackage{epstopdf}
\usepackage{amsmath,amssymb,graphicx}
\usepackage{mathrsfs}  
\usepackage{latexsym}
\usepackage{amssymb}

\newcommand{\C}{\mathbb{C}}
\newcommand{\CP}{\mathbb{CP}}

\newcommand{\RP}{\mathbb{RP}}
\newcommand{\R}{\mathbb{R}}

\newcommand{\Z}{\mathbb{Z}}

\newcommand{\PP}{\mathbb{P}}

\newcommand{\PA}{\mathbb{PA}}

\newcommand{\OO}{\cal{O}}

\newcommand{\Rho}{\mathrm{P}}

\def\ep{{\varepsilon}}

\def\ov{\overline}

\def\p{\partial}

\def\g{\mathfrak{g}}

\newcommand{\bea}{\begin{eqnarray}}
\newcommand{\eea}{\end{eqnarray}}

\def\be{\begin{equation}}
\def\ee{\end{equation}}

\def\v{\bf v}
\def\u{\bf u}
\newcommand{\e}{\textbf{e}}

\def\O{\cal O}
\def\OO{\cal O}
\def\om{\omega}

\def\ov{\overline}
\def\g{\mathfrak{g}}

\def\tw{\tilde w}
\def\tz{\tilde z}

\def\p{\partial}

\newcommand{\spp}{\mathbb{S}}
\def\ov{\overline}

\newcommand{\hook}{{\setlength{\unitlength}{11pt}   
                   \begin{picture}(.833,.8)
                   \put(.15,.08){\line(1,0){.35}}
                   \put(.5,.08){\line(0,1){.5}}
                   \end{picture}}}

\def\l{\lambda}

\def\O{{\cal O}}

\newtheorem{theo}{Theorem}[section]

\begin{document}

\title{
{\bf Twistor theory at fifty: from contour integrals to twistor strings}
\vskip 20pt}

\author{Michael Atiyah\thanks{m.atiyah@ed.ac.uk}
\\[1 pt]
{\sl \small School of Mathematics, University of Edinburgh,}  \\[-3pt] {\sl \small King’s Buildings,  Edinburgh EH9 3JZ, UK. }\\
{\sl \small and}\\
{\sl \small Trinity College, Cambridge, CB2 1TQ, UK.}
 \\[10pt]
Maciej Dunajski\thanks{m.dunajski@damtp.cam.ac.uk}\\[1pt]
{\sl \small Department of Applied Mathematics and Theoretical Physics,}\\[-3pt] {\sl \small University of Cambridge, Wilberforce Road, Cambridge CB3 0WA, UK.} 
\\[10pt]
Lionel J Mason\thanks{lmason@maths.ox.ac.uk}
\\[5pt] {\sl \small The Mathematical Institute, Andrew Wiles Building,}\\[-3pt] {\sl\small University of Oxford, ROQ Woodstock Rd, OX2 6GG}}
\date{September 6, 2017} 
\maketitle
\begin{center}
{\small Dedicated to Roger Penrose and Nick Woodhouse at 85 and 67 years.}
\end{center}

\begin{abstract}
We review aspects of twistor theory, its aims and achievements spanning the  
last five decades.  In the twistor approach, space--time is secondary with events being derived objects that correspond to compact holomorphic curves in 
a complex three--fold -- the twistor space.  
After giving an elementary construction of this space we demonstrate how 
solutions to linear and nonlinear equations of mathematical physics:
anti-self-duality (ASD) equations on Yang--Mills, or conformal curvature  can be encoded into twistor cohomology. These twistor correspondences yield
explicit examples of Yang--Mills, and gravitational instantons which 
we review. They also underlie the twistor approach to integrability:
the solitonic systems arise as symmetry reductions of ASD Yang--Mills equations, and Einstein--Weyl dispersionless systems are reductions of ASD 
conformal equations.  

 We then review the holomorphic string theories in twistor and ambitwistor 
spaces, and explain how these theories give rise to  remarkable 
new formulae for the computation of quantum scattering amplitudes.
Finally we discuss the Newtonian limit of twistor theory, and its possible
role in Penrose's proposal for a role of gravity
in quantum collapse of a wave function.
\end{abstract}


\thispagestyle{empty}

{\small \tableofcontents}
\newpage




\maketitle

\section{Twistor Theory}
\label{introduction}
Twistor theory was originally proposed as a new geometric framework for physics that aims to unify
general relativity and quantum mechanics \cite{Penrose_twistor_alg, penrose_2,
PR86, penrose_mc, penrose_ward}. In the twistor approach, space--time is secondary with events being derived objects that correspond to compact holomorphic curves in a complex three--fold, the \emph{twistor space}.
The mathematics of twistor
theory goes back  to the 19th century Klein correspondence, but we shall begin
our discussion with  a formula for solutions
to the wave equation in (3+1)--dimensional Minkowski space--time  put forward by  Bateman in 1904 \cite{Bateman}
\be
\label{contour_form}
\phi(x,y,z,t)=\oint_{\Gamma\subset \CP^1}
f((z+ct)+(x+i y)\lambda, (x-i y) 
-(z-ct)\lambda, \lambda)d\lambda.
\ee
This is  the most elementary of Penrose's series of twistor integral formulae for massless fields  
\cite{Pen_2}.
The closed contour $\Gamma\subset \CP^1$ encloses some poles
of a meromorphic function $f$.
Differentiating  (\ref{contour_form}) under the integral sign yields 
\be
\label{wave_eq}
\frac{1}{c^2}\frac{\p^2 \phi}{\p t^2}-\frac{\p^2 \phi}{\p x^2}-
\frac{\p^2 \phi}{\p y^2}-\frac{\p^2 \phi}{\p z^2}=0.
\ee
The twistor contour integral formula (\ref{contour_form}) is a paradigm for how twistor theory should work and is a good starting point for discussing its development over the last five decades. In particular one may ask
\begin{itemize}
\item {\em What does this formula mean geometrically?}

The integrand of (\ref{contour_form}) is a function of three complex arguments and we will see in
\S{\ref{sect_twistor_space}} that these arise as local affine coordinates on 
projective twistor space $\mathbb{PT}$ which we take to be
$\CP^3-\CP^1$. In  (\ref{contour_form}) the coordinates on $\mathbb{PT}$ are 
restricted to a  line with affine coordinate $\lambda$. 
The Minkowski space arises as a real slice in the four-dimensional space of lines in $\mathbb{PT}$. 

The map (\ref{contour_form}) from functions $f$ to solutions to the wave equation is not one to one:  functions holomorphic inside
$\Gamma$ can be added to $f$ without changing the solution $\phi$. This freedom in $f$ was understood in the 1970s in a fruitful interaction between the Geometry and Mathematical Physics research groups in Oxford  \cite{michael_personal}: twistor functions such as $f$ in
(\ref{contour_form})
 should  be regarded as elements
of $\check{\mathrm C}$ech sheaf cohomology groups. Rigorous theorems establishing twistor correspondences
for the wave equation, and higher spin linear equations have now been established
\cite{EPW, WW89, Hitchin_linearfields, BE}. The concrete realisations of these theorems lead to (contour) integral formulae.

\item {\em Do `similar' formulae exist for nonlinear equations of mathematical physics, such as 
Einstein or Yang--Mills equations?}

The more general integral formulae of Penrose \cite{Pen_2} give solutions
to both linearised 
Einstein and Yang--Mills equations.   In the case that the linearised field is anti-self-dual (i.e., circularly polarised or right handed) these cohomology classes correspond to linearised deformations of the complex structure of twistor space for gravity \cite{Pe76, AHS} or of a vector  bundle in the Yang-Mills case \cite{Wa77}. 
We shall review these constructions in \S \ref{sect_gravity} 
and \S{\ref{sect_gauge}}.

These constructions give an  {\em `in principle'} general solution to the equations in the sense that locally every solution can be represented locally in terms of free data on the twistor space as in the original integral formula.  Indeed this   leads to large
classes of explicit examples (e.g.\ Yang-Mills and  gravitational instantons which we
shall  review in the gravitational case 
in \S\ref{sect_gravity}\ref{sect_grav_inst}) although it can be hard to implement for  
general solutions. 

It turns out \cite {Wa85} that most known integrable systems arise as symmetry reductions of 
either the anti--self--dual Yang--Mills or the anti--self--dual 
(conformal) gravity equations. The twistor constructions then reduce to known (inverse scattering transform, dressing method, \dots) or new solution generation techniques for soliton
and other integrable equations \cite{MW98, Dbook}. We shall review some of this development in 
\S\ref{sect_integrable}.

As far as the full Einstein and Yang-Mills equations are concerned, the situation is less satisfactory.  The generic nonlinear  fields can be encoded  in terms of complex geometry in closely related ambitwistor spaces.  In these situations the expressions of the field equations are less straightforward and  they no longer seem to provide a general solution generation method.  Nevertheless, they have still had major impact on the understanding of these theories in the context of perturbative quantum field theory as we will see in \S\ref{sect_twistor_strings}. 

\item {\em Does it all lead to interesting mathematics?}

 The impacts on mathematics have been an unexpected major spin--off from the original twistor 
programme. These range over  geometry in the study of hyper--K\"ahler manifolds \cite{AHS,H79,K1,K2,salamon},
 conformal, CR  and projective structures 
\cite{GJMS,BEG,sparling_CR, bryant_2, claude_2, sparling_penrose, claude_2,
ambient, cap_gover, BDE, para_con_BE,E_S,baum,West}, exotic 
holonomy \cite{bryant_exotic,mer,mer2,mer3}, in
representation theory \cite{BE,parabook} and differential equations particularly in the form of integrable systems \cite{MW98,Dbook}.  We will make more specific comments and references in the rest of this review.

\item {\em Is it physics?}

Thus far, the effort has been to reformulate conventional physics in twistor space rather than propose new theories.  It has been hard to give a complete reformulation of conventional physics on twistor space in the form of nonlinear  generalizations of (\ref{contour_form}).  Nevertheless, in just the past 13  years, holomorphic string theories in twistor and ambitwistor spaces have provided twistorial  formulations of a full range of theories that are commonly considered in particle physics.  They also provide remarkable new formulae for the computation of scattering amplitudes.  Many technical issues that remain to be resolved to give a complete reformulation of conventional physics ideas even in this context of peturbative quantum field theory.  Like conventional string theories, these theories do not, for example, have a satisfactory non-perturbative definition.  Furthermore, despite recent advances at one and two loops, their applicability to all loop orders has yet to be demonstrated.     See 
\S\ref{sect_twistor_strings} for a full discussion.

The full (non-anti--self--dual) Einstein and Yang--Mills equations are not integrable
and so one does not expect a holomorphic twistor description of their solutions that has the simplicity of their integrable self-dual sectors. 
It is hoped that the full, non--perturbative implementation of twistor theory in physics is still to be revealed. One set of ideas builds on Penrose's proposal for a role of gravity
in quantum collapse of a wave function \cite{Pe96,Pe11}. This proposal only makes use of Newtonian gravity, but it is the case that in the Newtonian limit the self--dual/anti--self--dual constraint
disappears from twistor theory and all physics can be incorporated
in the $c\rightarrow \infty$ limit of ${\mathbb{PT}}$ \cite{DG16}, see 
\S\ref{sec_newton}.

\item{\em Does it generalise to higher dimensions?}

There are by now many  generalisations of twistors in dimensions higher than four \cite{salamon,hughston2, hughston, hugston3,murray,sorokin,
berk,arman,Delduc:1991ir,Galperin:1992pz,Uvarov:2007vs}.   One definition takes twistor space to be the projective pure spinors of the conformal group.  This definition respects full conformal invariance, and there are analogues of  (\ref{contour_form}) for massless fields.  However,  the (holomorphic) dimension of such twistor
spaces goes up quadratically in dimension and become higher than  the dimension of the Cauchy data (i.e., one less than the dimension of
space--time). Thus  solutions to the wave equation and its non--linear generalisations do not map to  unconstrained twistor data and this is also reflected in the higher degree of the cohomology classes in higher dimensions that encode massless fields.  These do not seem to have straightforward nonlinear extensions. 

Another  dimension agnostic generalisation of twistor theory is via \emph{ambitwistors}.  Indeed some of the ambitwistor string models described in \S\ref{sect_twistor_strings} are only critical in 10 dimensions, relating closely to conventional string theory, although without the higher massive modes. 

Twistor theory has many higher dimensional analogues for space-times of restricted  holonomy \cite{salamon}.  The hyper-K\"ahler case of manifolds of dimension $4k$ with holonomy in $SU(2)\times SP(2k)$  admit a particularly direct generalisation of Penrose's original nonlinear graviton construction and now has wide application across mathematics and physics. 
\end{itemize}

This review celebrates the fifty years of twistor theory since the publication of the first paper on the subject\footnote{See also the programme and slides from the meeting, New Horizons in Twistor Theory in Oxford January 2017 that celebrated this anniversary along with the 85th birthday or Roger Penrose and the 67th of Nick Woodhouse. http://www.maths.ox.ac.uk/groups/mathematical-physics/events/twistors50.}  by Roger Penrose \cite{Penrose_twistor_alg}. 
We apologize to the many researchers whose valuable contributions have been inadvertently  overlooked. 

\section{Twistor space and incidence relation}
\label{sect_twistor_space}
Twistor theory is particularly effective in dimension four because of an interplay between three isomorphisms. Let $M$ be a real oriented four--dimensional manifold with a metric
$g$ of arbitrary signature.
\begin{itemize}
\item
The Hodge $\ast$ operator is an
involution on two-forms, and induces a decomposition
\be 
\label{splitting}
\Lambda^{2}(T^*M) = \Lambda_{+}^{2}(T^*M) \oplus \Lambda_{-}^{2}(T^*M)
\ee
of two-forms into self-dual (SD)
and anti-self-dual (ASD)  components, which
only depends on the conformal class of $g$. 
\item
Locally there exist complex rank-two vector bundles $\spp, \spp'$  (spin-bundles) over $M$ equipped with parallel symplectic structures
$\ep, \ep'$ such that
\be
\label{can_bun_iso}
T_\C M\cong {\spp}\otimes {\spp'}
\ee
is a  canonical bundle isomorphism, and
\be
\label{metric_abc}
g(p_1\otimes q_1,p_2\otimes q_2)
=\varepsilon(p_1,p_2)\varepsilon'(q_1, q_2)
\ee
for $p_1, p_2\in \Gamma(\spp)$ and $q_1, q_2\in \Gamma(\spp')$.
The isomorphism (\ref{can_bun_iso}) is related to (\ref{splitting}) by
\[
\Lambda_{+}^{2}\cong {\spp'}^*\odot {\spp'}^*, \quad
\Lambda_{-}^{2}\cong {\spp}^*\odot {\spp}^*.
\]
\item
The orthogonal group in dimension four is not simple: 
\be
\label{group_iso}
SO(4, \C)\cong (SL(2, \C)\times \widetilde{SL}(2, \C))/\Z_2
\ee
where $\spp$ and $\spp'$ defined above are the representation spaces
of $SL(2)$ and $\widetilde{SL}(2)$ respectively.
There exist three real slices: In the Lorentzian signature $Spin(3, 1)
\cong SL(2, \C)$ and both copies
of $SL(2, \C)$ in (\ref{group_iso}) are related by complex conjugation. In the Riemannian signature 
$Spin(4, 0)= SU(2)\times \widetilde{SU}(2)$. In $(2, 2)$ (also called neutral, or ultra-hyperbolic
signature) $Spin(2, 2)\cong SL(2, \R)\times \widetilde{SL}(2, \R)$. Only in this signature
there exists a notion of real spinors, and as we shall see in \S\ref{22reality} real twistors.
\end{itemize}
\subsection{Incidence relation}
The projective twistor space ${\mathbb{PT}}$ is defined to be
$\CP^3-\CP^1$.  The homogeneous coordinates of a twistor
are $(Z^0, Z^1, Z^2, Z^3)\sim (\rho Z^0, \rho Z^1, \rho Z^2, \rho Z^3)$,  where $\rho\in \C^*$ 
and $(Z^2, Z^3)\neq (0, 0)$. 
The projective twistor space (which we shall call twistor space from now on) 
and  Minkowski space are linked by
the incidence relation
\be
\label{4d_incidence}
\left (
\begin{array}{cc}
Z^0\\
Z^1 
\end{array}
\right )
=
\frac{i}{\sqrt{2}}
\left (
\begin{array}{cc}
ct+z & x+iy\\
x-iy & ct-z
\end{array}
\right )
\left (
\begin{array}{cc}
Z^2 \\
Z^3 
\end{array}
\right )
\ee
where $x^\mu=(ct, x, y, z)$ are coordinates of a point in Minkowski space.
If two points in  Minkowski space
are incident with the same twistor, then they are connected by a null line.
Let
\[
\Sigma(Z, \ov{Z})=Z^0\ov{Z^2}+Z^1\ov{Z^3}
+Z^2\ov{Z^0}+Z^3\ov{Z^1}
\]
be a $(+ + - -)$  Hermitian inner product
on the non--projective twistor space ${\mathbb{T}}=\C^4-\C^2$. The orientation--preserving endomorphisms of 
the twistor space which preserve $\Sigma$ form a group $SU(2, 2)$ which is locally isomorphic to the conformal group $SO(4, 2)$ of Minkowski space. The twistor space ${\mathbb{T}}$ is divided  into three parts
depending on whether $\Sigma$ is positive, negative or zero. This partition descends
to the projective twistor space. In particular
the hypersurface
\[
{\cal PN}=\{[Z]\in {\mathbb{PT}}, \Sigma(Z, \ov{Z})=0\}\subset {\mathbb{PT}}
\]
is preserved by the conformal transformations of the Minkowski space
which can be verified directly using (\ref{4d_incidence}).
The five dimensional manifold ${\cal PN}\cong S^2\times \R^3$
is the space of light rays in the Minkowski space.
Fixing the coordinates $x^\mu$ of a  space--time point 
in (\ref{4d_incidence}) gives a plane in the non--projective twistor
space $\C^4-\C^2$ or a projective line $\CP^1$ in ${\mathbb{PT}}$.
If the coordinates $x^\mu$ are real this line lies in the
hypersurface ${\cal PN}$. Conversely, fixing a twistor
in ${\cal PN}$ gives a light--ray in the Minkowski space.

So far only the null twistors (points in ${\cal PN}$) have been 
relevant in this discussion. General points in ${\mathbb{PT}}$
can be interpreted in terms of the {\em complexified Minkowski space}
$M_\C=\C^4$ where they correspond to  $\alpha$--planes, i. e. null two--dimensional planes with
self--dual tangent bi-vector. This, again, is a direct consequence of
(\ref{4d_incidence}) where now the coordinates $x^{\mu}$ are complex:
\begin{center}
\includegraphics[width=7cm,height=3cm,angle=0]{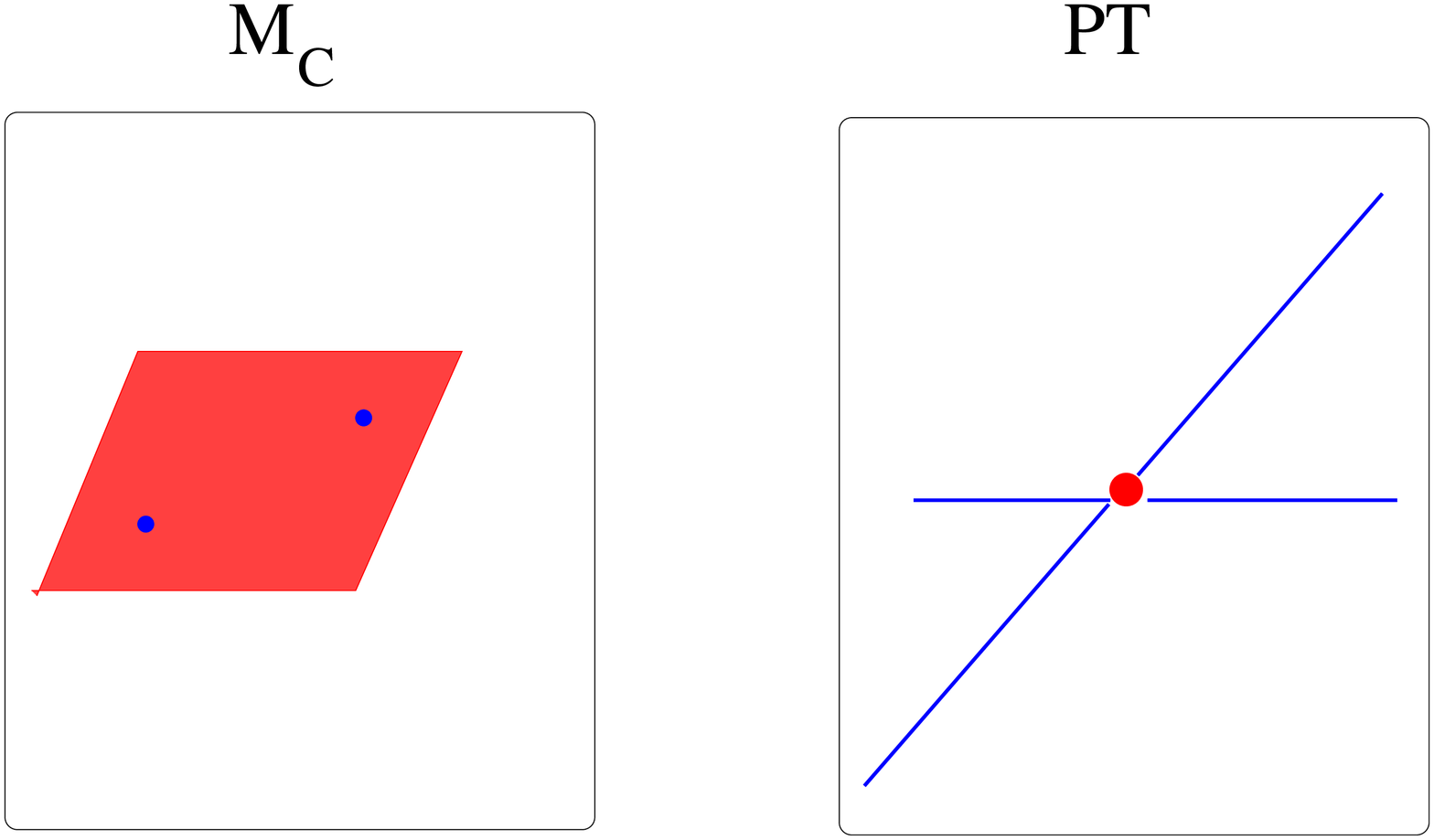}
\begin{center}
{{\bf Figure 1.} {\em Twistor incidence relation}}
\end{center}
\end{center}
{ \begin{eqnarray*}
\mbox{Complexified space-time $M_\C$}\quad&\longleftrightarrow&\mbox{Twistor space ${\mathbb{PT}}$}\\
\mbox{Point $p$}\quad&\longleftrightarrow&\mbox{Complex line}\,\, \
L_p=\CP^1\\
\mbox{Null self-dual (=$\alpha$) two-plane }\quad&\longleftrightarrow&\mbox{Point}.\\
p_1, p_2 \;\mbox{null separated}  \quad&\longleftrightarrow&
L_1, L_2 \; \mbox{intersect at one point.}
\end{eqnarray*}
\subsection{Robinson Congruence}
The non--null twistors can also be interpreted in the real 
Minkowski space, but this is somewhat less obvious \cite{Penrose_twistor_alg}: The  
inner product
$\Sigma$ defines a vector space ${\mathbb{T}}^*$ dual to the 
non--projective twistor space. Dual twistors are 
the elements of the projective space ${\mathbb{PT}}^*$.
Consider a twistor $Z\in {\mathbb{PT}}\setminus{\cal PN}$. Its dual $\ov{Z}\in{\mathbb{PT}}^*$ corresponds to a 
two--dimensional complex
projective plane $\CP^2$ in ${\mathbb{PT}}$. This holomorphic plane intersects the space of light rays
${\cal PN}$ in a real three--dimensional locus corresponding to
a three--parameter family of light--rays in the real Minkowski space. 
The family of light rays
representing a non--null twistor is called the Robinson congruence.
\begin{center}
\includegraphics[width=7cm,height=3cm,angle=0]{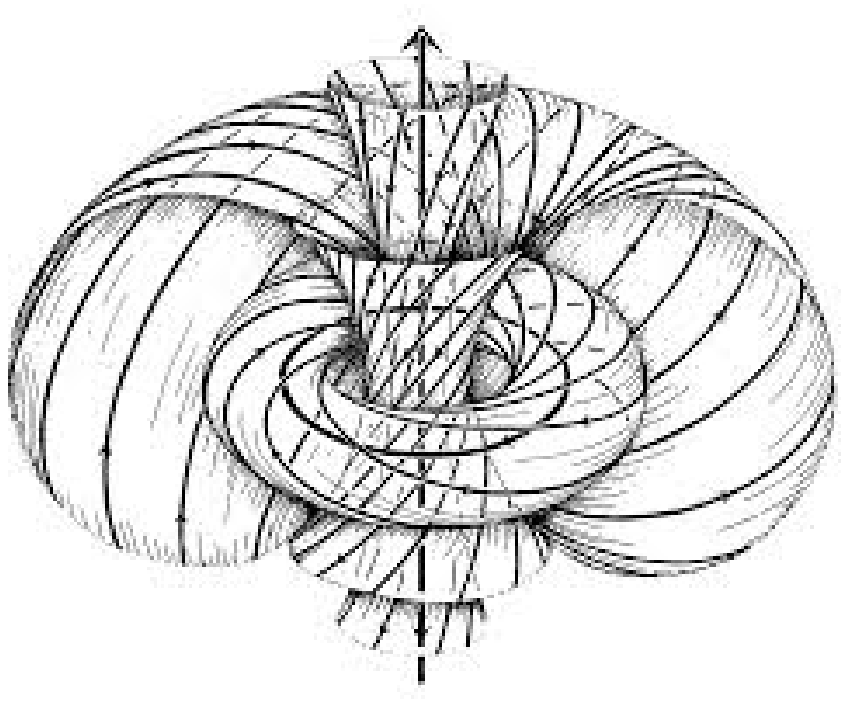}
\begin{center}
{{\bf Figure 2.} {\em Robinson congruence of twisting light rays.}}
\end{center}
\end{center}

\begin{eqnarray*}
\mbox{Null ray in $M$}\quad&\longleftrightarrow&\mbox{
Point in ${\cal PN}$}\\
\mbox{Robinson congruence 
$\{\ov{Z}\}\cap{\cal PN}$  }\quad&\longleftrightarrow&\mbox{Point in }\,\, \  {\mathbb{PT}} \setminus {\cal PN}. 
\end{eqnarray*}

The Robinson congruence in Figure 2 
is taken from the front cover of \cite{PR86}.
It consists of a system of twisted oriented circles in $\R^3$:
a light--ray is represented by a point in $\R^3$ together with an arrow 
indicating
the direction of the ray's motion. 
It is this twisting property of circles in Figure 2 which
gave rise to a term `twistor' for points of ${\mathbb{PT}}$.
An account of congruences in general relativity which motivated initial progress in twistor theory is given in \cite{kerr, Rob, Rob_T}.
\subsection{Cohomology}
The twistor interpretation
of Penrose's contour integral formula (\ref{contour_form}) is as follows.
Cover the twistor space ${\mathbb{T}}=\C^4-\C^2$ by two open sets: $U_0$ 
defined 
by $Z^2\neq 0$ and $U_1$ defined by  $Z^3\neq 0$.
Consider a function on the non--projective twistor space
$f=f(Z^0, Z^1, Z^2, Z^3)$  which is holomorphic on $U_0\cap U_1$
and homogeneous of degree $-2$ in $Z^{\alpha}$. Restrict this function to a two--dimensional plane
in ${\mathbb{T}}$ defined by the incidence relation (\ref{4d_incidence}) with $(x, y, z, t)$ fixed.
This gives rise
to an element of the cohomology group\footnote{The cohomology group $H^1(\CP^1,{\OO}{(k)})$  is the space of 
functions $f_{01}$ holomorphic on $U_0\cap U_1$ and homogeneous of degree
$k$ in coordinates $[Z^2, Z^3]$ modulo addition  of coboundaries (functions holomorphic
on $U_0$ and $U_1$). In a trivialisation over $U_0$ we represent  $f_{01}$ by a holomorphic function 
$f$ on $\C^*$. In the trivialisation over $U_1$, $f_{01}$ is represented
by $\lambda^{-k}f$, where $\lambda=Z^3/Z^2$. Here, and in the rest of this paper ${\cal O}(k)$ denotes a line bundle over $\CP^1$
with transition function $\lambda^{-k}$ in a trivialisation over $U_0$. Alternatively it is defined as the
$(-k)$th tensor power of the tautological line bundle.} 
$H^1(L_p, {\cal O}(-2))$
on the projective twistor space, where   $L_p\cong \CP^1$ is the curve corresponding, via the incidence relation  (\ref{4d_incidence}), to a point $p\in M_\C$. Integrate the cohomology class  along 
a contour $\Gamma$ in $L_p$. This gives (\ref{contour_form}) with 
$\lambda =Z^3/Z^2$.   For example  $f=(P_\alpha Z^{\alpha})^{-1}(Q_\beta Z^{\beta})^{-1}$,
where $\alpha, \beta= 0, \dots, 3$ and $(P_\alpha, Q_\beta)$ are
constant dual twistors gives rise to a fundamental solution to the wave equation (\ref{wave_eq}).
The Theorem of \cite{EPW} states that solutions to the wave equation
(\ref{wave_eq})  which holomorphically extend to a future tube domain in $M_\C$ are in one--to--one
correspondence with elements of the cohomology group $H^1({\mathbb{PT}}, {\cal O}(-2))$. This correspondence
extents to solutions of zero--rest--mass equations with higher spin, and elements
of $H^1({\mathbb{PT}}, {\cal O}(k))$ where $k$ is an integer.
 See \cite{EPW, Hitchin_linearfields, PR86, WW89, HT} for further details.
\section{Twistors for curved spaces}
\label{sect_gravity} 
The twistor space of complexified Minkowski space $M_\C=\C^4$ was defined by the incidence relation
(\ref{4d_incidence}) as the space of all $\alpha$--planes in $\C^4$. 
Let $(M_\C, g)$ be a 
holomorphic four--manifold with a holomorphic Riemannian
metric and a holomorphic volume form.  Define an $\alpha$--surface to be a two--dimensional surface in 
$M_\C$ such that its tangent plane at every point is an $\alpha$--plane.
If the metric $g$ is curved, there will be integrability conditions coming from the Frobenius Theorem  for an $\alpha$--plane to be tangent to a two--dimensional surface.
\subsection{The Nonlinear Graviton construction}
Define ${\mathbb{PT}}$  to be the  space of $\alpha$--surfaces $\zeta$ in  $(M_\C, g)$. The Frobenius theorem implies that for 
$X, Y\in T\zeta\rightarrow [X,  Y]\in
T\zeta$, and there are obstruction in terms of the curvature of $g$.
This gives rise to  the Nonlinear Graviton Theorem 
\begin{theo}[Penrose \cite{Pe76}]
\label{penrose} 
There exists a three--parameter family of $\alpha$--surfaces in $M_\C$ iff the the Weyl tensor of $g$  is anti--self--dual, i.e. 
\be
\label{self_duality}
C_{abcd}=-\frac{1}{2}{\epsilon_{ab}}^{pq}C_{cdpq}.
\ee
\end{theo}
The anti--self--duality of the Weyl tensor is the property of the 
whole conformal class 
\[
[g]={\Omega^2g, \quad \Omega:M_\C\rightarrow \C^*}
\]
rather than any particular metric. Points in an ASD conformal
manifold $(M_\C, [g])$ correspond to rational curves
in ${\mathbb{PT}}$ with normal bundle ${\OO}(1)\oplus{\OO}(1)$, and points
in ${\mathbb{PT}}$ correspond to $\alpha$--surfaces in $M_\C$. The ASD
conformal structure on $M_\C$ can be defined in terms of algebraic
geometry of curves in twistor space: ${\mathbb{PT}}$ is three dimensional, so two curves in ${\mathbb{PT}}$ generically do not
intersect. Two points in $M_\C$ are null separated if and only if
the corresponding curves in  ${\mathbb{PT}}$ intersect at one point.
\begin{theo}[Penrose \cite{Pe76}]
\label{theo3}
Let $M_\C$ be a moduli space of all rational curves with the normal bundle
${\OO}(1)\oplus{\OO}(1)$ in some complex three--fold ${\mathbb{PT}}$. Then 
$M_\C$ is a complex four--fold with a holomorphic conformal metric
with anti--self--dual curvature. Locally all ASD holomorphic conformal metrics
arise from some ${\mathbb{PT}}$.
\end{theo}
More conditions need to be imposed on ${\mathbb{PT}}$ if the conformal structure contains a Ricci-flat metric. In this case there exists a holomorphic fibration $\mu:{\mathbb{PT}}\rightarrow\CP^1$ 
with ${\cal O}(2)$-valued symplectic form on the fibres. Other curvature
conditions (ASD Einstein \cite{Ward_cosmo,Hi82,lebrun_h,vandoren,hoegner}, Hyper--Hermitian \cite{joyce, D99}, 
scalar--flat K\"ahler \cite{pontecorvo}, null K\"ahler \cite{D02})
can also be encoded in terms of additional  holomorphic structures
on ${\mathbb{PT}}$. Some early motivation
for Theorem \ref{theo3} came from complex general relativity, and theory
of ${\cal H}$--spaces. See \cite{Newman,pleban}.
\subsection{Reality conditions}
\label{22reality}
The real ASD conformal structures are obtained by introducing
an involution on the twistor space. If the conformal structure
has Lorentizian signature, then the anti--self--duality implies
vanishing of the Weyl tensor, and thus $g$ is conformally flat.
This leaves  two possibilities:
 Riemannian and neutral signatures. In both cases the involutions act on the  twistor lines, 
thus giving rise to maps from $\CP^1$ to $\CP^1$:
the antipodal map which in stereographic
coordinates is given by $\l\rightarrow -1/\ov{\lambda}$, or
a complex conjugation which swaps the lower and upper hemispheres
preserving the real equator. The antipodal map has no fixed
points and
corresponds to the positive--definite conformal structures.
The conjugation corresponds to the neutral case.

In the discussion below we shall make use of the double fibration picture
\be
\label{doublefib}
{M_\C}\stackrel{r}\longleftarrow 
{\cal F}\stackrel{q}\longrightarrow {\mathbb{PT}},
\ee
where the five--complex--dimensional 
correspondence is defined by
\[
{\cal F}={\mathbb{PT}}\times {M_\C}|_{\zeta\in L_p}= {M_\C}\times\CP^1
\] 
where $L_p$ is the line in ${\mathbb{PT}}$ that corresponds to 
$p\in {M_\C}$ and $\zeta\in{\mathbb{PT}}$ lies on $L_p$. The space
${\cal F}$ can be identified with a projectivisation $\PP{\spp'}$ of the spin bundle 
$\spp'\rightarrow M_\C$. It is equipped with a rank-2 distribution, the {\em twistor distribution},
which at a given point $(p, \lambda)$ of ${\cal F}$ is spanned by horizontal lifts of vectors spanning
$\alpha$--surface at $p\in M_\C$. The normal bundle to
$L_p$ consists of vectors tangent to $p$ horizontally lifted to
$T_{(p,\lambda)}{\cal F}$ modulo the twistor distribution ${\cal D}$. 
We have a sequence of sheaves over $\CP^1$
\[
0\longrightarrow {\cal D} \longrightarrow \C^4 \longrightarrow
{\cal O}(1)\oplus{\cal O}(1)\longrightarrow 0.
\]
Using the abstract index notation \cite{PR86}
(so that, for example,  $\pi^{A'}$ denotes a section of $\spp'$, and
no choice of a local frame or coordinates is assumed)
the map $\C^4 \longrightarrow {\cal O}(1)\oplus{\cal O}(1)$ is given by
$V^{AA'}\longrightarrow V^{AA'}\pi_{A'}$.  Its kernel consists of
vectors of the form $\pi^{A'}\rho^A$ with $\rho\in \spp$ varying. The
twistor distribution is therefore ${\cal D}={\cal O}(-1)\otimes \spp$ and so there
is a canonical $L_A\in\Gamma(D\otimes {\cal O}(1)\otimes \spp)$, given by $L_A=\pi^{A'}\nabla_{AA'}$,  where $A=0, 1$.
\begin{itemize}
\item{\bf Euclidean case.}
The conjugation 
$\sigma:\spp'\rightarrow \spp'$ given by 
 $\sigma(\pi_{0'}, \pi_{1'})=(\ov{\pi_{1'}},
-\ov{\pi_{0'}})$
descends from  $\spp'$ to an involution $\sigma:{\mathbb{PT}}\rightarrow {\mathbb{PT}}$
such that $\sigma^2=-{\mbox{Id}}$. The twistor curves
which are preserved by $\sigma$ form a four--real parameter
family, thus giving rise to a real four--manifold $M_\R$. If 
$\zeta\in {\mathbb{PT}}$
then $\zeta$ and $\sigma(\zeta)$ are connected by  a unique {\em real curve}.
The real curves do not intersect as no two points are connected by
a null geodesic in the positive definite case. Therefore there exists a fibration of the 
twistor space ${\mathbb{PT}}$ over a real four--manifold $M_\R$. A fibre over a point $p\in M_\R$
is a copy of a $\CP^1$. The fibration is not holomorphic, but
smooth. 

In the Atiyah--Hitchin--Singer \cite{AHS} version of the correspondence
the twistor space of the positive definite metric
is a real six--dimensional manifold
identified with the projective spin bundle
$P(\spp')\rightarrow M_{\R}$. 

Given a conformal structure $[g]$ on $M_{\R}$ 
one defines an almost--complex--structure on $P(\spp')$ by declaring
\[\{ \pi^{A'}\nabla_{AA'},  {\p}/{\p\ov{\lambda}}\}\]
to be the anti--holomorphic vector fields in $T^{0, 1}(P(\spp'))$. 

\begin{theo}[Atiyah--Hitchin--Singer \cite{AHS}]
\label{theo4}
The six--dimensional almost--complex manifold
\[
P(\spp')\rightarrow M_\R
\] 
parametrises almost--complex--structures in 
$(M_\R, [g])$. Moreover $P(\spp')$ is complex iff $[g]$ is ASD.
\end{theo}
\item{\bf Neutral case.}
The spinor conjugation  
$\sigma:\spp'\rightarrow \spp'$ given by 
 $\sigma(\pi_{0'}, \pi_{1'})=(\ov{\pi_{0'}}, \ov{\pi_{1'}})$
allows an
invariant decomposition of a spinor 
into its real and 
imaginary part, and  thus definition 
of  {\em real $\alpha$-surfaces} \cite{woodhouse_john, D02}.


In general $\pi=\mbox{Re}{(\pi)}+i \mbox{Im}{(\pi)}$, and 
the correspondence 
space ${\cal F}=P(\spp')$  decomposes into two open sets 
\begin{eqnarray*}
{\cal F}_+&=&\{ (p, [\pi])\in {\cal F};
\mbox{Re}({\pi_{A'}})\mbox{Im}({\pi^{A'}})>0\}={ M_\R}\times D_+,\\
{\cal F}_-&=&\{ (p, [\pi])\in {\cal F};
\mbox{Re}({\pi_{A'}})\mbox{Im}({\pi^{A'}})<0\}={ M_\R}\times  D_-,
\end{eqnarray*}
where $D_{\pm}$ are two copies of a Poincare disc.
These sets are separated by a real correspondence space
${\cal F}_0={M_\R}\times\RP^1.$
The correspondence  spaces ${\cal F}_{\pm}$ have the structure 
of a complex manifold
in a way similar to the AHS Euclidean picture. 
There exists an $\RP^1$ worth of real $\alpha$--surfaces through each point in $M_\R$, 
and real twistor distribution consisting of vectors tangent 
to real $\alpha$--surfaces defines a foliation of ${\cal F}_0$
with  quotient ${\mathbb{PT}}_0$ which leads to 
a double fibration:
\[
{M_\R}\stackrel{r}\longleftarrow 
{\cal F}_0\stackrel{q}\longrightarrow {\mathbb{PT}}_0.
\]
The twistor space  ${\mathbb{PT}}$ is a union of two open subsets
${\mathbb{PT}}_+=({\cal F}_+)$ and ${\mathbb{PT}}_-=({\cal F}_-)$
separated by a three-dimensional real boundary ${\mathbb{PT}}_0$.


These reality conditions are relevant in the twistor approach to integrable systems (see \S\ref{sect_integrable}), integral geometry, twistor inspired computations of scattering amplitudes  (see \S\ref{sect_twistor_strings}), as well as recent applications \cite{Anew0} of the Index Theorem \cite{AS} which do not rely on positivity of the metric. 
The discussion in this subsection has assumed real analyticity
of $M_\R$. The approach of LeBrun and Mason \cite{LM} based on 
holomorphic discs  can weaken this assumption.
\end{itemize}
\subsection{Kodaira Deformation Theory}
One way of obtaining complex three--manifolds
with four--parameter families of ${\OO}(1)\oplus{\OO}(1)$ curves comes from the Kodaira deformation theory applied
to ${\mathbb{PT}}=\CP^3-\CP^1$
\begin{center}
\includegraphics[width=5cm,height=3cm,angle=0]{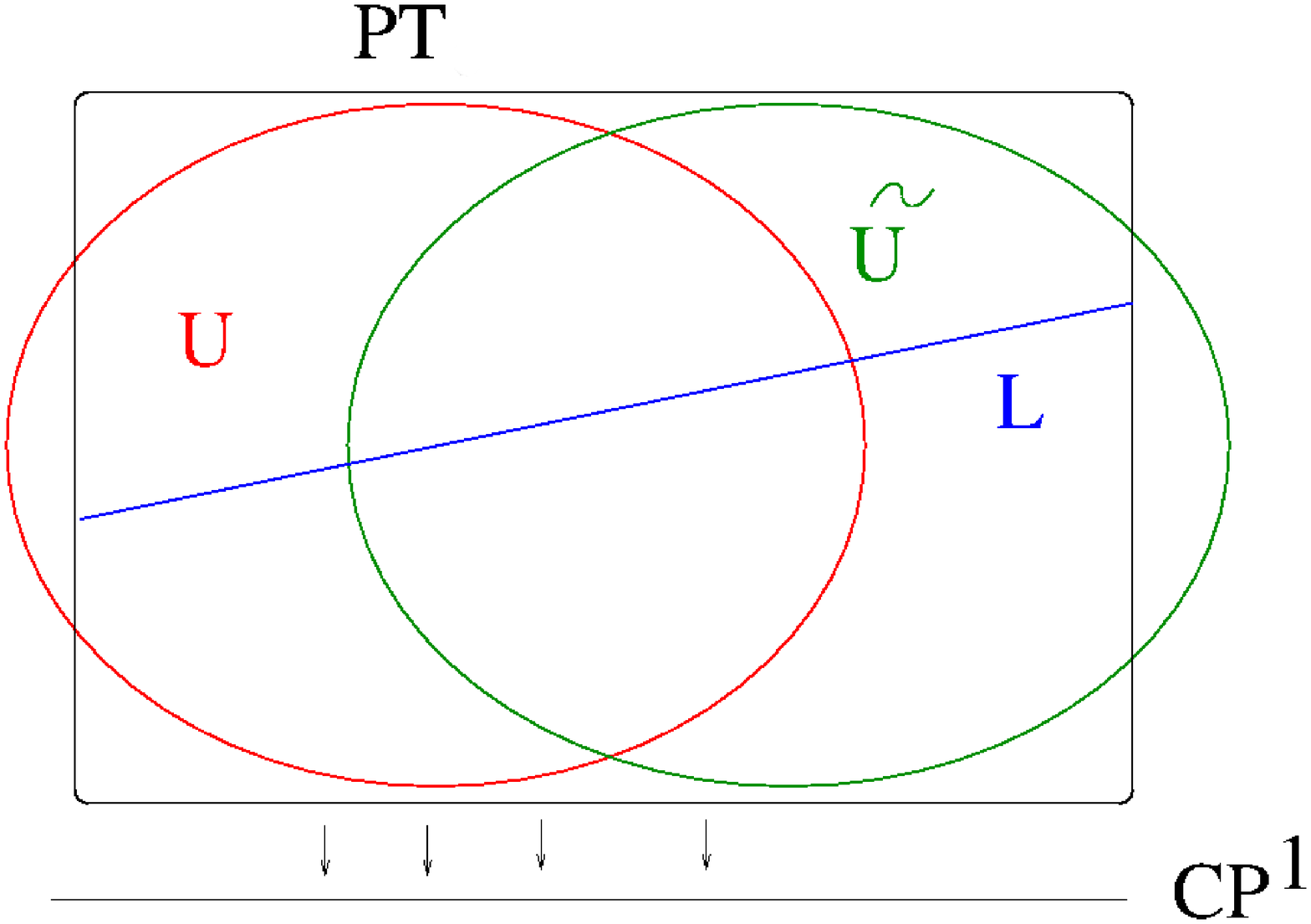}\quad
\includegraphics[width=5cm,height=3cm,angle=0]{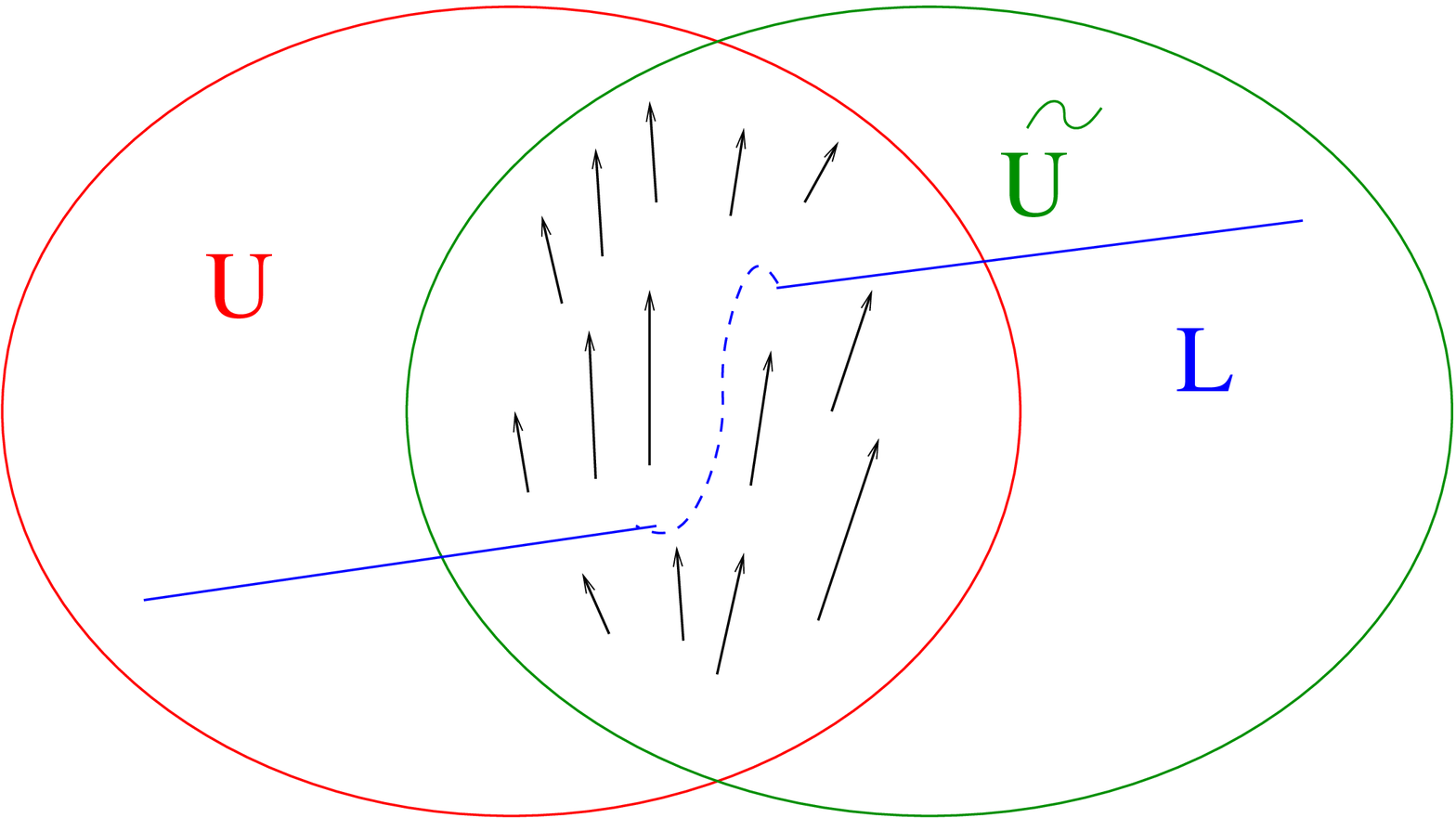}
\begin{center}
{{\bf Figure 3.} {\em Curvature on $(M_\C, g)$ corresponds to 
deformations of ${\mathbb{PT}}$}}
\end{center}
\end{center}
The normal bundle $N(L_p)\equiv T(PT)|_{L_p}/TL_p\cong {\OO}(1)\oplus{\OO}(1)$
satisfies \[
H^1(L_p, N(L_p))=0.\]
The Kodaira theorems \cite{Kodaira} imply that there exist infinitesimal deformations
of the complex structure of ${\mathbb{PT}}$ which preserve the four parameter family $M_\C$ of $\CP^1$s,
as well as the type of their normal bundle. Moreover this deformed family admits an isomorphism
\[
H^0(L_p, N(L_p))\cong T_p M_\C
\]
identifying tangent vectors to $M_\C$ with pairs of linear homogeneous polynomials in two variables.
This identification allows to construct a conformal structure on $M_\C$ arising from a quadratic condition that both polynomials in each pair have a common zero. There are some examples
of ASD Ricci flat metrics arising from explicit deformations - see \cite{HT, Dbook}. A method of constructing such examples was pioneered by George Sparling.
\subsubsection{Twistor solution to the holonomy problem.}
The Kodaira approach to twistor theory has given rise
to a complete classification of manifolds with exotic holonomy groups 
(holonomy groups of affine connections which are missing from Berger's list).
The first landmark step was taken by Robert Bryant \cite{bryant_exotic} 
who generalised
the Kodaira theorems and the twistor correspondence to Legendrian curves. Complex contact three--folds with 4--parameter
family of Legendrian rational curves with normal bundle
${\OO}(2)\oplus{\OO}(2)$ correspond to four manifolds $M_\C$
such that $T_pM_\C\cong \C^2\odot\C^2\odot\C^2$ and there exists
a torsion--free connection with holonomy group $GL(2, \C)$.
The theory was extended by Merkulov to allow Legendrian deformations
of more  general submanifolds \cite{mer,mer2}. This work lead to a complete
classification by Merkulov and Schwachhofer \cite{mer3}.
\subsection{Gravitational Instantons}
\label{sect_grav_inst}
Gravitational instantons are solutions to the Einstein equations in Riemannian signature which give complete metrics whose  curvature is  concentrated in a finite region of a space-time. The non--compact gravitational instantons asymptotically 
`look like' flat space. While not all gravitational instantons
are (anti)--self--dual (e.g. the Euclidean Schwarzchild solution is not)
most of them are, and therefore they arise from Theorems
\ref{theo3} and \ref{theo4}.
\begin{itemize}
\item There exists a large class of gravitational instantons
which depend on a harmonic function on $\R^3$: 
\be
\label{GH_metric}
g=V(dx^2+dy^2+dz^2)+V^{-1}(d\tau+ A)^2, \quad \mbox{where}\quad 
*_3dV=dA.
\ee
where $V$ and $A$ are
a function and a one--form respectively which do not depend
on $\tau$.  This is known as the 
Gibbons-Hawking  ansatz \cite{GH}. The resulting metrics are hyper--K\"ahler
(or equivalently anti--self--dual and Ricci flat).
The Killing vector  field  $K=\p/\p \tau$ is tri--holomorphic - it
preserves the sphere of K\"ahler forms of $g$. It gives rise
to a holomorphic vector field on the corresponding twistor space
which preserves the ${\cal O}(2)$--valued symplectic structure on the fibres
of ${\mathbb{PT}}\rightarrow \CP^1$. Therefore there exists an associated
${\cal O}(2)$ valued Hamiltonian, and the Gibbons--Hawking twistor
space admits a global fibration over the total space
of ${\cal O}(2)$. Conversely, any twistor space
which admits such fibration leads to the Gibbons--Hawking
metric on the moduli space of twistor curves \cite{TW79, HKLR87}.
\item An example of a harmonic function in (\ref{GH_metric})
which leads to the Eguchi--Hanson gravitational instanton is
$V=|{\bf r}+{\bf a}|^{-1} +|{\bf r}-{\bf a}|^{-1}$. It is 
asymptotically  locally Euclidean (ALE), as it approaches 
$\R^4/\Z^2$ for large $|{\bf r}|$. The corresponding twistor space
has been constructed by Hitchin \cite{H79}.
\item A general gravitational instanton  
is called ALE if it approaches $\R^4/\Gamma$ at infinity, where
$\Gamma$ is a discrete subgroup of $SU(2)$. Kronheimer \cite{K1, K2}
has constructed ALE spaces for finite subgroups 
\[A_k, D_k, E_6, E_7, E_8\] 
of $SU(2)$. In each case the twistor space
is a three--dimensional hyper--surface
\[
F(X, Y, Z, \lambda)=0
\]
in the rank three bundle 
${\cal O}(p)\oplus {\cal O}(q) \oplus {\cal O}(r)\rightarrow \CP^1$,
for some integers $(p,q,r)$, 
where $F$ is a singularity resolution of one of the Klein polynomials
corresponding to the Platonic solids
\[
XY-Z^k=0, X^2+Y^2Z+Z^k=0, \dots, X^2+Y^3+Z^5=0 \quad \mbox{(isocahedron.)}
\]
The twistor spaces of these ALE instantons admit a holomorphic
fibration over the total space of ${\cal O}{(2n)}$ for some $n\geq 1$. In case of $A_k$ one has $n=1$ and the metric belongs to the Gibbons--Hawking class. In the remaining cases $n>1$, and the 
resulting metrics do not admit any tri--holomporphic Killing vector. They do however admit hidden symmetries 
(in the form of tri--holomorphic Killing spinors), and arise from 
a generalised Legendre transform \cite{LR1,DM,Bielawski,boyer}.
\item There are other types of gravitational instantons which are not ALE, and are characterised
by different volume growths of a ball of the given geodesic radius 
\cite{ck,hein}. They are
ALF (asymptotically locally flat), and `inductively' named  ALG, ALH spaces. 
Some ALF spaces arise from the Gibbons--Hawking ansatz 
(\ref{GH_metric})
where 
\[
V = 1 + \sum_{m=1}^N \frac{1}{\mid {\bf x} - {\bf x}_m \mid}
\]
where ${\bf x}_1, \dots, {\bf x}_N$  are fixed points in $\R^3$
(the corresponding twistor spaces are known), but others do not. In 
\cite{ch}
some progress has been made in constructing twistor spaces for $D_k$ ALF instantons, but
finding the twistor spaces, or explicit local forms for the remaining cases is a open problem.
\end{itemize}
There also exist compact examples of Riemannian metrics with ASD conformal curvature.
The round $S^4$ and $\CP^2$ with the Fubini--Study metric are explicit 
examples where the ASD metric
is also Einstein with positive Ricci scalar. A Ricci--flat ASD metric is known to exist
on the $K3$ surface, but the explicit formula for the metric is not known.

LeBrun has proven \cite{L91} that there are ASD metrics with positive scalar curvature
on any connected sum $N\overline{\CP}^2$ of reversed oriented complex projective planes. This class, together with a round four--sphere exhaust all simply connected possibilities. The corresponding twistor spaces can be constructed
in an algebraic way.
The strongest result belongs to Taubes \cite{taubes}. If $M$ is any compact oriented smooth
four--manifold, then there exists some $N_0>0$ such that 
\[
M_N=M\# N \overline{\CP}^2
\]
admits an ASD metric for any $N\geq N_0$.
\section{Local Twistors}
\label{sec_loca}
There exists at least three definitions of a twistor which agree in a four--dimensional flat space.
The first,  twistors as $\alpha$--planes,  was used in the last section, where its curved generalisation lead to the Nonlinear Graviton construction and anti--self--duality. The second,
twistors as spinors for the conformal group, relies heavily on maximal symmetry  and so does not generalise to curved metrics. 
The last, twistors as solutions to the twistor equation, leads to interesting notions
of a local twistor bundle and a local 
twistor transport \cite{penrose_mc,Dighton,PR86,sparling_nature} which we now review.
 
We shall make use of the isomorphism (\ref{can_bun_iso}).
Let $Z^{\alpha}$ be homogeneous coordinates of a twistor
as in \S\ref{sect_twistor_space}.  Set $Z^{\alpha}=(\omega^{A}, \pi_{A'})$.
Differentiating the  incidence relation (\ref{4d_incidence}) yields
\be
\label{twistor_eq}
{\nabla_{A'}}^{(A}\omega^{B)}=0,
\ee
where ${\nabla_{A'}}^{A}=\epsilon^{AB}\nabla_{BA'}$, and $\epsilon^{AB}=
-\epsilon^{BA}$ is a (chosen) symplectic forom on $\spp^*$ used to raise and 
lower indices.

The space--time coordinates $(x, y, z, t)$  are constants of integration resulting from solving this equation on $M_\C$. Let us consider (\ref{twistor_eq}) on a general curved four--manifold, where it is called {\em the twistor equation}. It is conformally invariant under the transormations of the metric 
$g\rightarrow \Omega^2 g$.
The prolongation of the twistor equation leads to a connection 
on a rank--four vector bundle $\spp\otimes{\cal E}[1]\oplus \spp'$ called the {\em local twistor bundle}.
Here ${\cal E}[k]$ denotes a line bundle of conformal densities of weight $k$.
This connection also called the {\em local twistor transport}, and is given 
by \cite{Dighton}
\[
{\cal D}_a 
\left(\begin{array}{c}
\omega^B\\ 
\pi_{B'}
\end{array} \right)= 
\left(\begin{array}{c} \nabla_a\omega^{B}-{\epsilon_{A}}^B\pi_{A'} \\ 
\nabla_a\pi_{B'}-\Rho_{ABA'B'}\omega^B
\end{array} \right),
\]
where $\Rho_{ab}$ is the Schouten tensor of conformal geometry given by
\[
\Rho_{ab}=\frac{1}{2}R_{ab}-\frac{1}{12}Rg_{ab}.
\]
The holonomy of the local twistor transport obstructs existence of global twistors on curved four manifolds (all local normal forms of Lorentzian metrics admitting solutions to (\ref{twistor_eq})
have been found in \cite{lewandowski}).

The tractor bundle is isomorphic to the exterior square of the local twistor bundle. It is a rank--six
vector bundle ${\bf T}={\cal E}[1]\oplus T^*M\oplus{\cal E}[-1]$,
and its connection induced from the local twistor transform is
\be
\label{tractor_con}
{\quad{\mathcal{D}}_a \left(\begin{array}{c}
\sigma\\ 
\mu_b\\
\rho
\end{array} \right)= 
\left(\begin{array}{c} \nabla_a\sigma-\mu_a \\ 
\nabla_a\mu_b+\Rho_{ab}\sigma+g_{ab}\rho\\
\nabla_a\rho-{\Rho_a}^b\mu_b
\end{array} \right).}
\ee
This connection does not arise from a metric, but is related to a
pull back of the Levi--Civita connection of the so-called ambient metric
to a hypersurface. See \cite{ambient,cap_gover} as well
as \cite{sparling_CR} for discussion of the ambient construction.

The point about the connection (\ref{tractor_con}) is that it also arises
as a prolongation connection for the conformal to Einstein equation
\be
\label{conf_ein}
(\nabla_a\nabla_b+\Rho_{ab})_0\sigma=0,
\ee
where $(\dots)_0$ denotes the trace--free part.
If $\sigma$ satisfies (\ref{conf_ein}) where $\nabla_a$ and $\Rho_{ab}$ are computed from 
$g$, then $\sigma^{-2}g$ is Einstein \cite{lebrun_conf_einstein}.
Therefore the holonomy of (\ref{tractor_con}) leads to obstructions for 
an existence of an Einstein metric in a given conformal class 
\cite{BM,lebrun_conf_einstein,gover2,DTnew}. The Bach tensor is one of the obstructions
arising from a requirement that a parallel tractor needs to be annihilated
by the curvature of (\ref{tractor_con}) and its covariant derivatives.

Conformal geometry is a particular example of a parabolic geometry - a curved analog of a 
homogeneous space $G/P$ which is the quotient of a semi-simple Lie group $G$ by a parabolic subgroup $P$.
Other examples include projective, and CR geometries.
All parabolic geometries admit tractor connections. See \cite{parabook}
for details of these construction, and \cite{GJMS,gover,east_rice} where 
conformally  invariant differential operators  
 have been constructed. Examples of such operators are the twistor operator
$\omega_{A}\rightarrow \nabla_{A'(A}\omega_{B)}$ underlying  (\ref{twistor_eq}) and the operator
acting on ${\mbox{Sym}^4}(\spp^*)$
\[
C_{ABCD}\rightarrow (\nabla_{(A'}^C\nabla_{B')}^D+\Rho_{A'B'}^{CD})C_{ABCD}.
\]
This operator associates the conformally invariant Bach tensor to the anti--self--dual Weyl spinor.
\section{Gauge Theory}
\label{sect_gauge}
The full second--order Yang Mills equations on $\R^4$ are not integrable, and there is no twistor construction encoding their solutions in an unconstrained holomorphic data on ${\mathbb{PT}}$ - 
there do exist {\em ambitwistor constructions}
\cite{Witten_a, Yssenberg, Harnad} in terms of formal neighbourhods
of spaces of complex null geodesics, but they do not lead to any solution generation techniques.
As in the case of gravity, the anti--self--dual sub--sector can be described twistorialy, this time in terms of holomorphic vector
bundles over ${\mathbb{PT}}$ rather than deformations of its complex structures.
\subsection{ASDYM and the Ward Correspondence}
Let $A\in\Lambda^1(\R^4)\otimes\mathfrak{g}$, where
$\mathfrak{g}$ is some Lie algebra, and let
\[
F=dA+A\wedge A.
\]
The anti--self--dual Yang--Mills equations are
\be
\label{asdym}
*F=-F
\ee
where $*:\Lambda^2\rightarrow\Lambda^2$ is the Hodge endomorphism
depending on the flat metric and the orientation on $\R^4$.
These equations together with the Bianchi identity 
$DF:=dF+[A, F]=0$ imply the full Yang--Mills equations $D*F=0$.

Let us consider (\ref{asdym}) on the complexified Minkowski space
$M_\C=\C^4$ with a flat holomorphic metric and a holomorphic volume form. Equations (\ref{asdym}) are then equivalent to the vanishing of $F$ on each $\alpha$--plane in $M_\C$.
Therefore,  given $\zeta\in {\mathbb{PT}}$, there exists a vector space of solutions to
\be
\label{twistor_dis_1}
l^aD_a\Phi=0, \quad m^aD_a\Phi=0, \quad\mbox{where}\quad
\zeta=\mbox{span}\{l, m\}\in TM_\C.
\ee
The converse of this construction is also true, and leads
to a twistor correspondence for solutions to ASDYM equations
\begin{theo}[Ward \cite{Wa77}]
\label{main_twistor_theorem}
There is a one-to-one correspondence between:
\begin{enumerate}
\item Gauge equivalence classes of ASD connections on $M_\C$
with the gauge group $G=GL(n, \C)$,
\item Holomorphic rank--$n$ vector bundles ${E}$ over  twistor space
${{\mathbb{PT}}}$ which are trivial on each degree one section of 
${{\mathbb{PT}}}\rightarrow\CP^1$.
\end{enumerate}
\end{theo}
The splitting of the patching matrix $F_E$ for the bundle $E$ into a product of 
matrices holomorphic on $U_0$ and $U_1$ 
is the hardest part of this approach  to integrable PDEs.
When the Ward correspondence is reduced to lower dimensional PDEs  as in 
\S\ref{sect_integrable},
the splitting manifests itself as the Riemann--Hilbert 
problem in the dressing method.

 To obtain real solutions on $\R^4$ with the gauge group
$G=SU(n)$ the bundle must be compatible with the involution $\sigma$
preserving the Euclidean slice (compare \S\ref{sect_gravity}\ref{22reality}). 
This comes down to
$\det{F_E}=1$, and \[{F_E}^*(\zeta)=F_E(\sigma(\zeta)),\] 
where $*$ denotes the Hermitian conjugation,
and $\sigma:{\mathbb{PT}}\rightarrow {\mathbb{PT}}$  is the anti--holomorphic involution
on the twistor space which restricts to an antipodal map on each twistor line.  
See \cite{WW89,real_methods}.
\subsection{Lax pair}
Consider the complexified Minkowski space $M_\C=\C^4$ with 
coordinates $w, z, \tw, \tz$, and the metric and orientation
\[
g = 2(d z d\tz -d w d \tw), \quad 
\mbox{vol}=d w\wedge d \tw\wedge d  z\wedge d\tz.
\]
The Riemannian 
reality conditions are recovered if $\tilde{z}=\ov{z}, \tilde{w}=-\ov{w}$,
and the neutral signature arises if all four coordinates are taken to be real. 
The ASDYM equations (\ref{asdym})
arise as  the compatibility condition
for an overdetermined linear system $L\Psi=0,  M\Psi=0$, where
\be
\label{laxpair}
L=D_{\tz}-\l D_{w}, \qquad M=D_{\tw}-\l D_{z},
\ee
where 
$D_\mu=\p_{\mu}+[A_\mu, \cdot]$, 
and $\Psi=\Psi(w, z, \tw, \tz, \l)$ is the fundamental matrix solution.
Computing the commutator of the Lax pair $(L, M)$ yields
\[
[L, M]=F_{\tz\tw}-\l(F_{w\tw}-F_{z\tz})+\l^2 F_{wz}=0,
\]
and the vanishing of the coefficients of various powers of $\lambda$ gives
(\ref{asdym}). The geometric interpretation of this is as follows:
for each value of $\lambda\in \CP^1$ 
the vectors $l=\p_{\tz}-\l \p_{w},  m=\p_{\tw}-\l \p_{z}$ span a null plane in
$M_\C$ which is self--dual in the sense that 
$\om=\mbox{vol}(l, m, \dots, \dots)$ satisfies $*\om=\om$. The condition 
(\ref{asdym}) takes the equivalent form $\om\wedge F=0$,
thus $F$ vanishes on all $\alpha$ planes.
For a given YM potential $A$, the lax pair (\ref{laxpair}) can be expressed
as $L=l+l\hook A,  M=m+m\hook A$. 
\subsection{Instantons}
Instantons, i. e. solutions to ASDYM such that
\[
\int_{\R^4} \mbox{Tr}(F\wedge *F)<\infty
\]
extend from $\R^4$ to $S^4$.
The corresponding vector bundles extend from ${\mathbb{PT}}$ to $\CP^3$.
The holomorphic vector bundles over $\CP^3$ have
been extensively studied by algebraic geometers. 
All such bundles (and thus the instantons) can be generated by
the monad construction \cite{monads}.
\vskip5pt
One way to construct holomorphic vector bundles is to produce extensions of line bundles, which comes down to using upper-triangular matrices as patching functions.
Let $E$ be a rank--two holomorphic vector bundle over ${\mathbb{PT}}$
which arises as  an extension of a line bundle
$L\otimes\O(-k)$ by another
line bundle $L^*\otimes\O(k)$
\be
\label{exten}
0\longrightarrow L\otimes\O(-k)
\longrightarrow {E}\longrightarrow L^*\otimes\O(k)
\longrightarrow 0.
\ee
If $k>1$ then the YM potential
$A$ is given
in terms of a solution to the linear zero--rest--mass field equations with higher helicity.
\begin{theo}[Atiyah--Ward \cite{AW}]
Every $SU(2)$  ASDYM instanton 
over $\R^4$ arises from
a holomorphic vector bundle of the form (\ref{exten})
\end{theo}
Advances made in anti--self--dual gauge theory using the twistor methods lead to the studies of moduli spaces of connections. Such moduli spaces continue 
to play an important role in mathematical 
physics \cite{something}, and gave rise to major advances in the understanding of topology of four--manifolds \cite{donaldson}.
\subsection{Minitwistors and magnetic monopoles}
Another gauge theoretic problem which was solved using twistor methods 
\cite{H82, H83}
is the construction of non--abelian magnetic monopoles.

Let $(A, \phi)$  be a $\mathfrak{su}(n)$--valued one--form and a function respectively
on  $\R^3$, and let $F=dA+A\wedge A$. The non--abelian monopole equation is a system of non--linear PDEs
\be
\label{monopole_R3}
d\phi+[A, \phi]=*_3F
\ee
These are three equations for three unknowns as
$(A, \phi)$ are defined up 
to gauge transformations
\be
\label{gauge_trans_Ward_m}
A\longrightarrow gAg^{-1} -d g \; g^{-1}, \qquad
\phi\longrightarrow g\phi g^{-1}, \qquad g=g(x, y, z)\in SU(n)
\ee
and one component of $A$ can always be set to zero.

Following Hitchin \cite{H82} define the mini twistor space ${\cal Z}$ to be the space of oriented lines in $\R^3$. Any oriented line is of the form ${\v}+s{\u}, \quad s\in \R$
where $\u$ is a unit vector giving the direction of the line, and $\v$
is orthogonal to $\u$ and joints the line with some chosen point
(say the origin) in $\R^3$.
Thus
\[
{\cal Z}=\{({\u}, {\v})\in S^{2}\times\R^{3}, \;{\u.\v} =0 \}.
\]
For each fixed ${\u}\in S^2$ this space restricts to a tangent plane to $S^2$.  The twistor space is the union of  all tangent planes -- the tangent bundle $TS^{2}$ which is also a complex manifold
$T\CP^1$.
\begin{center}
\includegraphics[width=13cm,height=3cm,angle=0]{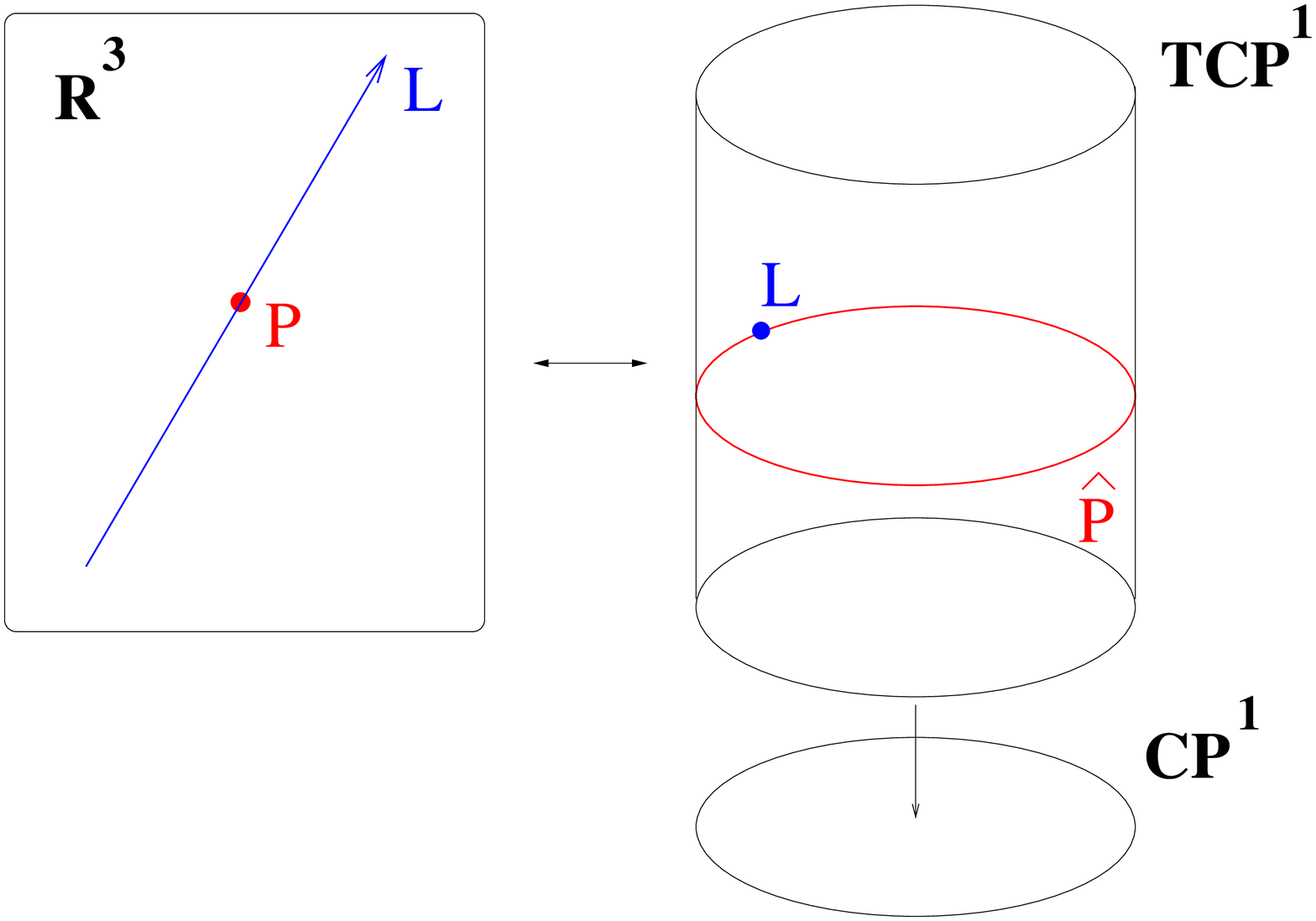}
\begin{center}
{{\bf Figure 4.} {\em Minitwistor Correspondence.}}
\end{center}
\end{center}
Given  $(A, \phi)$ solve a matrix ODE along each
oriented line ${\bf x}(s)={\bf v}+{s\bf u}$
\[
\frac{d V}{d s}+(u^jA_j+i\phi)V=0.
\]
This ODE assigns a complex vector space $\C^n$ to each point of ${\cal Z}$, 
thus giving rise to a 
complex vector bundle over the mini--twistor space.
Hitchin shows \cite{Hi82} that
monopole equation (\ref{monopole_R3}) on $\R^3$ holds if and only if
this vector bundle is holomorphic.

The mini--twistor space of Hitchin can also be obtained as a reduction of the twistor space
${\mathbb{PT}}=\CP^3-\CP^1$ by a holomorphic vector field corresponding to a translation in $\R^4$.
An analogous reduction of ASDYM on $\R^4$ by a rotation gives nonabelian hyperbolic monopoles \cite{A_monopole}.

In the next section we shall discuss how more general reductions of 
${\mathbb{PT}}$ give rise to solution generation techniques for lower dimensional integrable systems.

\section{Integrable Systems}
\label{sect_integrable}
Most lower dimensional integrable systems arise as symmetry reductions of anti--self--duality equations 
on $(M, [g])$ in  $(2, 2)$ or $(4 ,0)$ signature.

The solitonic integrable systems are reductions of ASDYM as their linear systems (Lax pairs) involve matrices. The program of reducing the ASDYM equations to various integrable equations has been proposed and initiated by  Ward \cite{Wa85} and fully implemented
in  the monograph \cite{MW98}.
The dispersionless integrable systems are reductions of anti--self--duality equations
on a conformal structure \cite{DMT00,Dbook}. A unified approach combining curved backgrounds with gauge theory has been developed by Calderbank \cite{Cal2}.

In both cases the reductions are implemented by assuming that the Yang--Mills potential
or the conformal metric are invariant with respect to a subgroup of the full group of conformal symmetries. Conformal Killing vectors on $M_\C$ correspond to holomorphic vector fields on ${\mathbb{PT}}$.
The resulting reduced system will admit a (reduced) Lax pair with a spectral  parameter 
coming from the twistor $\alpha$--plane distribution. It will be integrable by a reduced twistor correspondence of Theorem \ref{theo3} or Theorem \ref{main_twistor_theorem}.

\subsection{Solitonic equations}
The general scheme and classification of reductions 
of ASDYM on the complexified Minkowski space
involves a choice of subgroup  of the complex conformal group $PGL(4, \C)$, a real section (hyperbolic equations arise from ASDYM
in neutral signature), a gauge group and finally canonical forms of Higgs fields.

We have already seen one such symmetry reduction: ASDYM on $\R^4$ invariant under a one--dimensional group of translations 
generated by $K=\p/\p x^4$
reduce to the non--abelian monopole equation (\ref{monopole_R3}). The Higgs field
on $\R^3$ is related to the gauge potential $A$ on $\R^4$ by
$\phi=K\hook A$. The analogous reduction from $\R^{2, 2}$ leads to Ward's integrable chiral model 
on $\R^{2,  1}$ \cite{W88}. It is solved by a minitwistor construction, where the minitwistor space
${\cal Z}$ from the description of monopoles is instead equipped with an anti--holomorphic 
involution fixing a real equator on each twistor line \cite{W89}. The solitonic solutions
are singled out by bundles which extend to compactified mini--twistor spaces \cite{W98,prim}. Below we give some examples of reductions to two and one dimensions.
\begin{itemize}
\item Consider the $SU(2)$ ASDYM in neutral signature
and choose a gauge $A_{\tilde{z}}=0$. Let 
$T_{\alpha}, {\alpha}=1, 2, 3$ be two by two constant matrices such that
$
[T_{\alpha}, T_{\beta}]=-\epsilon_{\alpha\beta\gamma}
T_{\gamma}.
$
Then ASDYM equations are solved
by the ans\"atze
\[
A_w=2\cos{\phi}\,T_1+2\sin{\phi}\,T_2, \quad
A_{\tilde{w}}=2 T_1, \quad A_z= \p_z\phi\, T_3
\]
provided that $\phi=\phi(z, \tilde{z})$ satisfies 
\[
\phi_{z\tilde{z}}+4\sin{\phi}=0
\]
which is the Sine--Gordon equation. Analogous reductions
of ASDYM with gauge group $SL(3, \R)$ or $SU(2, 1)$ lead
to the Tzitzeica equations and other integrable systems arising in affine differential geometry
\cite{Dbook}. A general reduction by two translations on $\R^4$ lead to Hitchin's self--duality equations
which exhibit conformal invariance and thus extend to any Riemann surface \cite{He}.
\item Mason and Sparling \cite{MS} have shown that
any reduction to the ASDYM equations on $\R^{2, 2}$
with the gauge group  $SL(2, \R)$ by two translations exactly one of which is null is gauge equivalent
to either the KdV or the Nonlinear Schrodinger equation depending on whether the 
Higgs field corresponding to the null translation
is nilpotent or diagonalisable. In \cite{MS89} and \cite{MW98} this reduction has been extended to integrable hierarchies.
\item
By imposing three translational symmetries one can reduce 
ASDYM to an ODE.
Choose the Euclidean reality condition, and
assume that the YM potential is independent on $x^j=(x^1, x^2, x^3)$. 

Select  a gauge
$A_4=0$, and set $A_j=\Phi_j$,  where the Higgs fields 
$\Phi_j$ are real $\g$--valued 
functions of $x^4=t$.
The ASDYM equations reduce to the Nahm equations
\[
\dot{\Phi}_1=[\Phi_2, \Phi_3], \qquad \dot{\Phi}_2=[\Phi_3, \Phi_1], \qquad
\dot{\Phi}_3=[\Phi_1, \Phi_2].
\]
These equations admit a Lax representation which comes from taking a linear combination of $L$ and $M$ in (\ref{laxpair}).
Let
\[
B(\l)=(\Phi_1+i\Phi_2)+2\Phi_3\l-(\Phi_1-i\Phi_2)\l^2.
\]
Then
\begin{eqnarray*}
\dot{B} &=& [\Phi_2-i \Phi_1,\Phi_3] +2[\Phi_1,\Phi_2]\l-[\Phi_2+i \Phi_1,\Phi_3]\l^2\nonumber \\
 &=& [B,-i \Phi_3+i (\Phi_1-i \Phi_2)\l]\nonumber \\
 &=& [B, C], \qquad\mbox{where}\quad C= -i \Phi_3+i (\Phi_1-i \Phi_2)\l.
\end{eqnarray*}
The Nahm equations with the group of volume--preserving diffeomorphism
of some three-manifold as the gauge group are equivalent 
to ASD vacuum equations \cite{ash2}.
\item Reductions of ASDYM by three--dimensional abelian subgroups
of the complexifed conformal group $PGL(4, \C)$ lead to all six Painlev\'e equations \cite{MW98}. The coordinate--independent statement of the Painleve property for ASDYM was first put forward
by Ward \cite{Wa84}: If a
solution of ASDYM on $M_\C=\C^4$ has a non- characteristic singularity, then that singularity is at worst a pole.
Another twistor approach to the Painlev\'e 
equation is based on $SU(2)$--invariant anti--self--dual conformal structures \cite{tod_painleve,Hiso,maszczyk}.
\end{itemize}
\subsection{Dispersionless systems and Einstein--Weyl equations}
There is a class of integrable systems in 2+1 and three dimensions which do not fit into the framework described in the last
section. They do not arise from 
ASDYM and there is no 
finite--dimensional Riemann--Hilbert problem 
which leads to their solutions. These dispersionless integrable systems admit Lax representations which do not involve
matrices, like (\ref{laxpair}), but instead consist of vector fields.

Given a four--dimensional conformal structure $(M,[g])$ 
with a non-null conformal Killing  vector $K$, the three--dimensional space ${\cal W}$ of trajectories of $K$ inherits a conformal  structure $[h]$ represented by a metric
\[
h=g-\frac{K\otimes K}{|K^2|}.
\]
The ASD condition on $[g]$ results in an additional geometrical 
structure on $({\cal W}, [h])$; it becomes an {Einstein-Weyl} 
space \cite{JT}. There exists a torsion--free connection $D$ 
which preserves $[h]$ in the sense that
\be
\label{weyl_connection}
Dh=\omega\otimes h
\ee
for some one--form $\omega$, and the symmetrised Ricci tensor of $D$ is proportional to $h\in [h]$. These are the Einstein--Weyl 
equations \cite{cartanew}. They are 
conformally invariant: If
$h\longrightarrow \Omega^2 h$ then  $\omega\longrightarrow
\omega+2d{(\log{\Omega})}$.

Most known dispersionless integrable systems in $2+1$ and $3$
dimensions arise from the EW equations. Consult \cite{Dbook,Cal2,FK} for the complete list. Here we shall review the twistor picture, and examples of integrable reductions.
\begin{theo}[Hitchin \cite{Hi82}]
\label{theo_h}
There is a one--to--one correspondence between solutions to
Einstein--Weyl equations in three dimensions, and two--dimensional complex manifolds ${\cal Z}$ admitting a three parameter family of rational curves with normal bundle ${\OO}(2)$.
\end{theo}
\begin{center}
\includegraphics[width=8cm,height=3cm,angle=0]{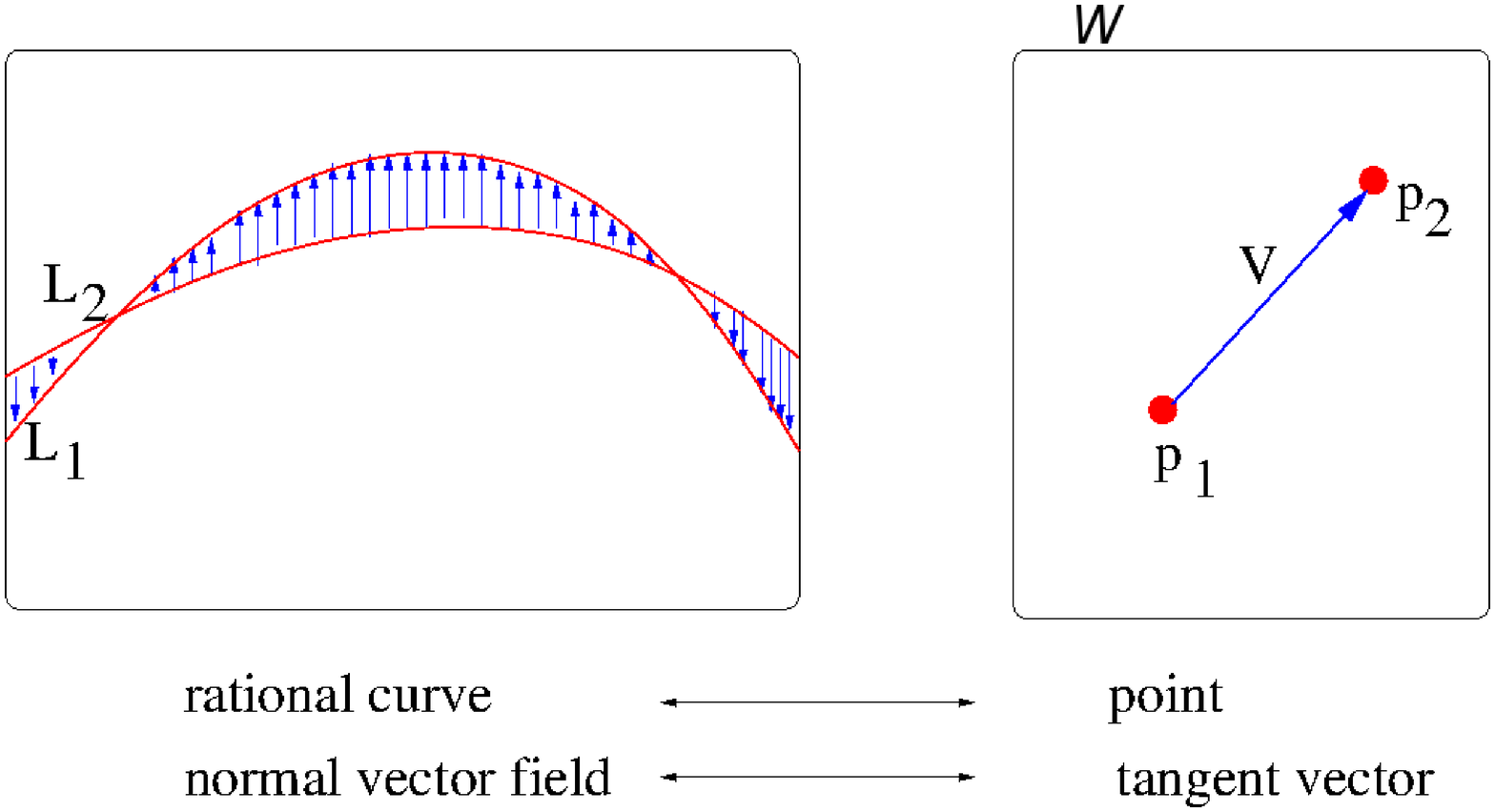}
\begin{center}
{{\bf Figure 5.} {\em Einstein--Weyl twistor correspondence.}}
\end{center}
\end{center}
In this twistor correspondence the points of ${\cal W}$ correspond 
to rational ${\OO}(2)$ curves in the complex surface ${\cal Z}$, and
points in  ${\cal Z}$ correspond to null surfaces
in ${\cal W}$ which are totally geodesic with respect to the connection $D$.

To construct the  conformal structure $[h]$ 
define  the null vectors at $p$ in ${\cal W}$ to be the
sections of the normal bundle $N(L_p)$ vanishing at some point
to second order. Any section of ${\OO}(2)$ 
is a quadratic polynomial, and the repeated root  condition is given
by the vanishing of its discriminant. This gives a quadratic
condition on $T{\cal W}$.

To define the connection
$D$, let a direction at $p\in {\cal W}$  be a one--dimensional
space of sections 
of ${\OO}(2)$  which vanishing at two points
$\zeta_1$ and $\zeta_2$ on a line $L_p$. The one--dimensional family of twistor 
${\OO}(2)$ curves
in ${\cal Z}$ passing through $\zeta_1$ and $\zeta_2$ gives a geodesic
 in ${\cal W}$ in a given direction. The limiting case
$\zeta_1=\zeta_2$ corresponds to   geodesics which 
are null with respect to $[h]$ in agreement with
(\ref{weyl_connection}). The special surfaces in ${\cal W}$ corresponding
to points in ${\cal Z}$ are
totally geodesic with respect to the connection $D$.
The integrability conditions for the existence of totally
geodesic surfaces is equivalent to the Einstein--Weyl equations \cite{cartanew}.

The dispersionless integrable systems can be encoded in the twistor correspondence of Theorem 
\ref{theo_h} if the twistor space admits some additional structures.
\begin{itemize}
\item If ${\cal Z}$ admits a preferred section of $\kappa^{-1/2}$,
where $\kappa$ is the canonical bundle of ${\cal Z}$, then there exist coordinates $(x, y, t)$ and a function $u$ on ${\cal W}$ such that 
\[
h=e^u(dx^2+dy^2)+dt^2, \quad \omega=2u_t dt
\]
and the EW equations reduce to the $SU(\infty)$ Toda 
equation \cite{ward_toda,L91}
\[
u_{xx}+u_{yy}+(e^u)_{tt}=0.
\]
This class of EW spaces admits both Riemannian and Lorentzian sections (for the later replace $t$ by $it$ or $x$ by $ix$),
which corresponds to two possible real structures on 
${\cal Z}$. It can be characterised on ${\cal W}$ by the existence
of twist--free shear--free geodesic congruence \cite{tod_ew, cal_petersen}.
\item If ${\cal Z}$ admits a preferred section of $\kappa^{-1/4}$,
then there exist coordinates $(x, y, t)$ and a function 
$u$ on ${\cal W}$ such that 
\be
\label{dkp_ew}
h=dy^2-4 dxdt-4udt^2,\quad \omega = -4u_xdt
\ee
and the EW equations reduce to the 
dispersion-less Kadomtsev-Petviashvili (dKP) equation \cite{DMT00}
\[
(u_t-uu_x)_x=u_{yy}.
\]
This class of EW spaces can be real only in the Lorentzian signature. The corresponding real structure on ${\cal Z}$
is an involution which fixes an equator on each $\CP^1$ and 
interchanges the upper and lower hemisphere. 
The vector  field $\p/\p x$ is null and covariantly constant with respect to the Weyl connection, and with weight $−1/2$. This vector field  gives rise to a parallel real weighted spinor, and finally to a preferred section of $\kappa^{-1/4}$. Conversely, any 
Einstein--Weyl structure which admits a covariantly constant weighted vector field is locally of the form (\ref{dkp_ew}) for some solution $u$  of the dKP equation.
\end{itemize}
The most general Lorentzian Einstein--Weyl structure corresponds \cite{DFK}
to the Manakov--Santini system \cite{Manakov_santini}. Manakov and Santini
have used a version of the non--linear Riemann--Hilbert problem and
their version of the inverse scattering transform to give
an analytical description of wave breaking in $2+1$ dimensions. It would be interesting to put their result in the twistor framework. The inverse scattering
transform of Manakov and Santini is intimately linked to the Nonlinear Graviton construction.  The coordinate form of the general conformal anti--self--duality
equation \cite{DFK} gives the master dispersionless integrable system in
$(2+2)$ dimensions, which is solvable by methods developed in  \cite{BDM,YS}.
\section{Twistors and scattering amplitudes}
\label{sect_twistor_strings}
Although there has been a longstanding programme to understand scattering amplitudes in twistor space via `twistor diagrams' \cite{penrose_mc,hodges_1}, the modern developments started with Witten's twistor-string \cite{Wi03} 
introduced in 2003.  The fallout has now spread in many directions. It  encompasses recursion relations that impact across quantum field theory but also back on the original twistor-diagram programme, Grassmannian integral formulae, polyhedral representations of amplitudes, twistor actions and ambitwistor--strings.
\subsection{Twistor-strings}
The twistor string story starts in the 1980's with a remarkable ampitude formula due to Parke and Taylor  \cite{parke_tay}, and its twistorial interpretation by Nair \cite{Nair:1988bq}.
Consider $n$ massles gluons, each carrying a null momentum  ${p_i}^{\mu}, i=1, \dots, n$.
The isomorphism (\ref{can_bun_iso}) and the fomula (\ref{metric_abc}) imply that null vectors are two-by-two matrices
with zero determinant, and thus rank one. Any such matrix is a tensor product of two spinors
\[
{p_i}^\mu={\pi_i}^{A'} {{\tilde{\pi}}_i} ^{A}\, .
\]
In spinor variables, the tree level amplitude for two negative helicity gluons and $n-2$ positive  leads
to \cite{parke_tay}
\be
\label{pt_amp}
{\cal A}_n=\frac{<\pi_i\pi_j>^4\delta^4(\sum_{k=1}^n p_k)}{<\pi_1\pi_2><\pi_2\pi_3>\cdots<\pi_{n-1}\pi_n><\pi_n\pi_1>},
\ee
where $ <\pi_k\pi_l>:=\epsilon_{A'B'}{\pi_k}^{A'}{\pi_l}^{B'}$, and
$i$th and $j$th particles are assumed to have negative helicity,
and the remaining particles have positive helicity.
Nair  \cite{Nair:1988bq} extended this formula to incorporate ${\cal N}=4$ supersymmetry and expressed it 
as an integral over the space of degree one 
curves (lines) in twistor space using a current algebra on each curve.

Witten  \cite{Wi03} extended this idea to provide a formulation of  ${\cal N}=4$ super Yang--Mills as a string theory whose target is the super--twistor space $\CP^{3|4}$ (see, e.g. \cite{Ferber:1977qx}).
This space has homogeneous coordinates $\mathcal Z^I=(Z^{\alpha},\chi^a)$ with $Z^\alpha$ the usual four bosonic homogeneous coordinates and $\chi^b$ four anti--commuting Grassmann coordinates $b=1,\ldots, 4$. The model is most simply described  \cite{Berkovits:2004hg} as a theory of holomorphic maps\footnote{These are D1 instantons in Witten's original B-model formulation.} $\mathcal{Z}:C\rightarrow \mathbb{T}$ from a closed Riemann surface $C$ to nonprojective twistor space.  It is based on the worldsheet action
\[
S=\int_C \mathcal{W}_I\bar{\partial}Z^I+ a\mathcal{W_I}\mathcal{Z}^I\, .
\]
Here $a$ is a $(0,1)$-form on the worldsheet $C$ that is a gauge field for  the scalings $(\mathcal{W},\mathcal{Z},a)\rightarrow (\e^{-\alpha}\mathcal{W}, \e^\alpha\mathcal{Z},a+\bar\partial \alpha)$. A prototype of this action was 
introduced in \cite{Shirafuji:1983zd}

 To  couple this to Yang--Mills we introduce a d-bar operator $\bar\partial +A$ on a region $\mathbb{PT}$ in super twistor space with $A$ a $(0,1)$-form taking values in some complex Lie algebra, and the field $J$ which is a $(1, 0)$ form
with values in the same Lie algebra on the worldsheet, and generates the current algebra\footnote{To turn this into a string theory one includes a chiral analogue of worldsheet gravity which when gauge is fixed just lead to ghosts and a BRST operator.}. When expanded out in  the fermionic variables $\chi$, such $A$ with $\bar{\partial}A=0$ give a Dolbeault cohomology classes in $H^1(\mathbb{PT},\mathcal O(p))$ for 
$p=0,\ldots, -4$ corresponding via the Penrose transform to the full multiplet for ${\cal N}=4$ super Yang-Mills from the positive helicty gluon to the fermions and scalars down to the negative helicity gluon.

The standard string prescription for amplitudes leads to a proposal for tree amplitudes for $n$ particles as correlators of $n$ `vertex operators' when $\Sigma=\CP^1$.  These take the form $V_i=\int_C {\mathrm{tr}} (A_i  J)$  
where $A_i$ is the `wave function', i.e., twistor cohomology class $H^1(\mathbb{PT},\mathcal{O})$, of the $i$th particle in the scattering, usually taken to be a (super-)momentum eigenstate and tr is  the Killing form on the Lie algebra.  
The correlators break up into contributions corresponding to the different degrees $d$ of the map from $C=\CP^1$ into twistor space.  The degree $d$ contribution gives the so-called N$^k$MHV amplitudes with $d=k+1$  where the MHV degree\footnote{MHV stands for maximal helicity violating because amplitudes vanish when $k<0$ or $k> n-2$.} $k$ corresponds to $k+2$ of the external particles having negative helicity with the rest positive.  

Let us give  a flavour of the formula.  If
$C$ is a degree-$d$ curve in twistor space, and ${\cal M}^d$ is the moduli space of curves containing $C$,
then, for any $(0, 1)$ form $A$ on $\CP^{3|4}$  with values
in $\mathfrak{gl}(N, \C)$, we can restrict $A$ to $C$ and consider $\overline{\p}_A|_C$.
The perturbative YM scattering amplitude is then obtained from the generating function 
\[
\int_{{\cal M}^d_{\R}}  \mbox{det}(\overline{\p}_{A}|_C) d\mu,
\]
where $d\mu$ is a holomorphic volume form on ${\cal M}^d$ and 
 ${\cal M}^d_{\R}$ a real slice  of ${\cal M}^d$. 
The role of $\det (\bar\partial_A |_C)$ is as the generating function of current algebra correlation functions on the curve $C$.  The $n$th variation of this functional with respect to $A$ as a $(0,1)$-form on $C$ gives the current algebra correlator that forms the denominator of the Parke--Taylor factor when $C$ is a line (together with some multitrace terms).  In the full formula, the $n$th variation with respect to $A$  with values in  the cohomology classes of linearized gauge fields for momentum eigenstates, is then the scattering amplitude as a function of these linear fields. 
In principle, the genus of $C$ determines the number of loops in the perturbative series, but the formulae have not been verified beyond genus 0.

This correlation function was evaluated for momentum eigenstates by Roiban, Spradlin and Volovich  (RSV) \cite{Roiban:2004yf} as a remarkably compact integral formula for the full S-matrix of ${\cal N}=4$ super Yang-Mills. 
The N$^k$MHV component is expressed as an integral over the moduli of rational curves of degree $k-1$ with $n$ marked points, a space of bosonic$|$fermionic dimension $(4d+n)|4(d+1)$.  Furthermore, many of the integrals are linear\footnote{ It is simplfied by the fact that the space of rational curves of degree $d$ in $\CP^{3|4}$ can be represented as homogeneous polynomials of degree $d$ in homogeneous coordinates, $\C^2$ for $\CP^1$ and so it is a vector space of dimension $4(d+1)|4(d+1)$ modulo $GL(2)$ which also acts on the marked points.} and can be performed explicitly reducing to a remarkable compact formula with $2d+2+n$ bosonic integrations against $2d+6+n|4(n-d-1)$
delta-functions (the excess bosonic delta functions expressing momentum conservation).  The resulting formula for the amplitude is, in effect, a sum of residues, remarkably simpler than anything that had been found before and difficult to imagine from the perspective of Feynman diagrams.

\subsection{Twistor-strings for gravity}
The Witten and Berkovits twistor string models also compute amplitudes of a certain conformal gravity theory with ${\cal N}=4$ supersymmetry \cite{berkow1}.  So it demonstrated the principle that gravity might be encoded in this way.  However, it is a problem for the construction of loop amplitudes for super Yang-Mills as these conformal gravity modes will of necessity run in the loops, although it is in any case still not clear whether the model can be used to calculate loop amplitudes even with conformal supergravity.  Furthermore, conformal gravity is widely regarded as a problematic theory, certainly quantum mechanically  because it necessarily contains negative norm states.

Although there was an early version of the Parke Taylor MHV formulae for gravity found in the 1980s \cite{Berends:1988zp}, the one given by Hodges  \cite{hodges_2} was the first to manifest permutation invariance and enough structure to suggest a version of  the RSV Yang-Mills formula    
for  maximally supersymmetric ${\cal N}=8$ gravity tree amplitudes.  This was discovered by Cachazo and Skinner 
\cite{Cachazo:2012kg}, see \cite{cms} for a proof and further developments. 
An underlying twistor string theory for this formula has been constructed by Skinner 
\cite{S13}, who showed that
${\cal N}=8$ supergravity is equivalent to string theory with target 
 $\CP^{3|8}$ but now with a supersymmetric worldsheet and some gauged symmetries. 


\subsection{The CHY formulae and ambitwistor   strings}
The twistor string theories of Witten, Berkovits, and Skinner gave a remarkable new paradigm for how twistor theory might encode genuine physics.  However, their construction very much relies on maximal supersymmetry and it is unclear how more general theories might be encoded.  The framework is also  
tied to four dimensions (for some this is a positive feature).  Although the string paradigm suggested that  multiloop processes should correspond to amplitudes built from higher genus Riemann surfaces, the details seem to be at best unclear and quite likely obstructed by anomalies.

In a parallel development, already before the Cachazo--Skinner formula for gravity amplitudes, Cachazo and coworkers had been exploring the relationship between the twistor-string amplitude formulae and a family of ideas, originating in conventional string theory, whereby gravity amplitudes can be expressed as the square of Yang-Mills amplitudes via the KLT relations \cite{Kawai:1985xq} and their extensions to colour-kinematic duality \cite{Bern:2010ue}.  This led to an independent twistor inspired formula for four dimensional gravity \cite{Cachazo:2012da}.  Cachazo, He and Yuan (CHY) subsequently refined and developed these ideas into a remarkably simple and elegant scheme for formulae analagous to those arising from the twistor-string for gauge theories and gravity (and a certain bi-adjoint scalar theory) in all dimensions \cite{Cachazo:2013iea}.  The framework has by now been extended to a variety of theories \cite{Cachazo:2014xea}.  In all of these formulae, the KLT idea of expressing gravity amplitudes as the square of Yang-Mills is essentially optimally realized (indeed much more elegantly than in KLT).  

The essential observation on which all these formulae rely is that the residues on which all the twistor-string formulae  of RSV  and so on are supported are configurations of points on the Riemann sphere  obtained from solutions to the \emph{scattering equations}.  These are equations for $n$ points on the Riemann sphere given the data of $n$ null momenta subject to momentum conservation that can be in any dimension.  They were first used  in \cite{Fairlie} to construct classical string solutions associated to $n$ particle scattering, but also arose from strings at strong coupling in calculations of Gross and Mende \cite{GM}.  On the support of the scattering equations, the complicated momentum kernel that forms the quadratic form in the gravity equals Yang-Mills squared in the KLT relations is diagonalized.

The question remained as to what the underlying string theory for these formulae might be.  This was answered in  \cite{MS14} based in its simplest form on a chiral, infinite tension limit of the ordinary bosonic string
\be
\label{ambi_action}
S=\int_C P\cdot {\overline{\p}} X-e P^2/2
\ee
where $(X, P)$ are coordinates on $T^*M$, and $P$ is understood to take values in 1-forms on the Riemann surface $C$. The Lagrange multiplier $e$ restricts the target
space to a hypersurface $P^2=0$ and is a gauge field for the action of the geodesic spray
$D_0=P\cdot\nabla$.  This is just a Lagrangian expression of the usual (holomorphic) symplectic reduction of the cotangent bundle to the space of (scaled) null geodesics $\mathbb{A} $.  In four dimensions this has become known as ambitwistor space as it is both the cotangent bundle of projective twistor space and of projective dual twistor and so chooses neither chirality.  However, it clearly exists in all dimensions as $\{T^*M|_{P^2=0}\}/D_0$, the space of complexified null geodesics. It can be defined
for geodesically convex regions in a complexification of any analytic Lorentzian or Riemannian manifold of any dimension. If $\mbox{dim}(M)=d$ then 
$\mbox{dim}(\PA)=2d-3$.

The study of ambitwistor space for $d=4$ started in 1978 with constructions by Witten and Isenberg, Yasskin and Green for Yang-Mills.  In four dimensions, ambitwistor space of   complexified Minkowski space $M_\C$ can be expressed as a quadric $Z^{\alpha}W_{\alpha}=0$ in $\mathbb{ PT}\times\mathbb{ PT}^*$. The ambitwistor  space is a complexification of the real hypersurface 
${\cal PN}\subset \mathbb {PT}$ introduced in 
\S\ref{sect_twistor_space}.
In \cite{Witten_a,Yssenberg,Harnad} (see also \cite{novikov2}) it was shown that generic analytic connections on bundles on Minkowski space correspond to topologically trivial bundles on ambitwistor space.  The full Yang-Mills equations can be characterised as the condition that the corresponding holomorphic 
vector bundles
on $\PA$ extend to  a third formal neighbourood in
 ${\mathbb{PT}}\times{\mathbb{PT}}^*$.  The Witten version \cite{Witten_a,Harnad} reformulated this third order extension condition to the simple requirement of the existence of the bundle on a supersymmetric extension of ambitwistor space built from  ${\cal N}=3$ supersymmetric twistor spaces. 
 This generalises the Ward correspondence 
(Theorem \ref{main_twistor_theorem}) but unlike this Theorem, it has not yet led to any effective solution generating techniques.

Gravitational analogues of the Witten, Isenberg, Yasskin and Green were developed by LeBrun, Baston and Mason 
\cite{BM,lebrun_conf_einstein}.
\begin{theo}[{LeBrun} \cite{lebrun_geod}]
The complex structure on $\PA$ determines the conformal structure
$(M_\C, [g])$. The correspondence is stable under deformations
of the complex structure on $\PA$ which preserve the contact 
structure $P\cdot dX$.

The  existence of $5^{th}$ order formal neighbourhoods corresponds to vanishing of the Bach tensor of the space-time and, when the space-time is algebraically general, a $6^{th}$ order extension corresponds to the space-time being conformally Einstein.
\end{theo}

The string theories based on the action (\ref{ambi_action}) have the property that they need to use ambitwistor cohomology classes arising from the ambitwistor Penrose transform in amplitude calculations; it is the explicit form of  these cohomology classes that imposes the scattering equations.  In order to obtain Yang-Mills and gravitational amplitudes from the theory, worldsheet supersymmetry needs to be introduced by analogy with the standard RNS superstring string and/or other worldsheet matter theories for other theories \cite{Casali:2015vta} (current algebras are required for the original biadjoint scalar and Yang-Mills formulae).

The ambitwistor string paradigm extends to many different geometrical realizations of ambitwistor space. Ambitwistor strings have also been constructed that are analogues of the Green-Schwarz string and the pure spinor string 
\cite{Bandos:2014lja, Berkovits:2013xba}.  In particular, the original twistor strings can be understood simply as arising from the four-dimensional realization of $\mathbb{A}$ as the cotangent bundle of projective twistor space. However, the approach of \cite{ivone} uses the  four dimensional realization of ambitwistor space as a subset of the product of twistor space with its dual to provide an ambidextrous twistorial ambitwistor string theory leading to new amplitude formulae for Yang-Mills and gravity that no longer depend on maximal supersymmetry. 

In another direction, one can realize ambitwistor space geometrically 
as the cotangent bundle $T^*\mathfrak{I}$ of null infinity $\mathfrak{I}$, \cite{Adamo:2014yya,Geyer:2014lca}.  This gives new insight into the relationship between asymptotic `BMS' symmetries and the soft  behaviour of scattering amplitudes, i.e., as the momentum of particles tends to zero.

An important feature of ambitwistor strings is that they lead to extensions of the CHY formulae to ones for loop amplitudes.
The original models of \cite{MS} are critical in 26 dimensions for the bosonic string and in 10 for the type II RNS version with two worldsheet supersymmetries.  This latter model was developed further 
in \cite{Adamo:2013tsa}, where it was shown that this theory does indeed correspond to full type II supergravity in 10 dimensions and  that anomalies all vanish for the computation of loop amplitudes.  An explicit conjecture for a formula at one loop for type II supergravity amplitudes was formulated 
using scattering equations on a torus in a scheme that in principle extends to all loop orders.  This formula was subsequently shown \cite{Geyer:2015bja} to be equivalent to one on a Riemann sphere with double points that does indeed compute amplitudes at one loop correctly.  Furthermore, on the Riemann sphere it is possible to see how to adapt the formulae to ones for many different theories in different dimensions with varying amounts of supersymmetry.  There remains the challenge to extend this to a scheme that transparently extends to all loop orders, but although the framework does work at two loops, it already there needs  new ideas \cite{Geyer:2016wjx}.
 
Much remains to be understood about how ambitwistor  strings reformulate  conventional massless theories, particularly at higher loops or in the nonlinear regime.  One key advance in the latter direction was the construction of ambitwistor strings on a curved background \cite{Adamo:2014wea} providing a kind of Lax pair for the full type II supergravity equations in 10 dimensions in the sense that the quantum consistency of the constraints is equivalent to the full nonlinear supergravity equations.

\subsection{Twistor actions}
At degree $d=0$ Witten argued that the effective field theory of the twistor-string is given by the twistor space action holomorphic Chern-Simons action 
\[
I=\int_{\CP^{3|4}}\Omega\wedge 
\mbox{Tr}(A\ov{\partial} A+
\frac{2}{3} A\wedge A\wedge A),
\]
where $\Omega=d^{3|4}Z$ is the natural holomorphic super-volume-form on $\CP^{3|4}$ (which turns out to be super-Calabi-Yau), and $A$ is a $(0,1)$-form. 
The field equations from this action are simply the integrability of the $\bar\partial$-operator, $\bar\partial_A^2=0$.  Interpreting this via the Ward transform (Theorem 
\ref{main_twistor_theorem}) on \emph{super}-twistor space now leads to 
the  spectrum a of full ${\cal N}=4$ super Yang--Mills, but with only anti-self-dual interactions.  

The question arises as to whether one can complete this action to provide the full interactions of super Yang-Mills.  This can be done by borrowing the leading part of the twistor-string computation and some experimentation shows that the interaction term
\[
S_{int}=\int_{{\cal M}^1_{\R}}  \log \, \mbox{det}(\overline{\p}_{A}|_C) d\mu,
\]
gives the correct interactions for full maximally supersymmetric Yang-Mills theory \cite{Mason:2005zm, Boels:2006ir}.  Here the integration is nonlocal in twistor space being firstly a nonlocal one over degree one rational curves (i.e., lines) and secondly over ${\cal M}^1_{\R}$ which is a real form
of the complexified Minkowski space - for example the Euclidean space.
This can be shown to be equivalent to the standard space-time action in Euclidean signature in a gauge that is harmonic on twistor lines as introduced in \cite{real_methods}. Despite its nonlocality, the resulting Feynman diagram formalism is remarkably tractable in an axial gauge in which a component of $A$ is set to zero in the direction of some fixed reference twistor $Z_*$.  This then leads \cite{bms} to the so-called MHV diagram formalism introduced informally in \cite{Cachazo:2004kj} by considering twistor-strings on degree-$d$ curves that arise as disjoint unions of $d$ lines.  In this formalism, amplitudes are computed via $1/p^2$ propagators and interaction vertices
 built from (\ref{pt_amp}) via a simple off-shell extension.  

The formulation of these rules via a twistor action allowed them to be expressed with maximal residual symmetry beyond the choice of $Z_*$, and extended from amplitudes to more general correlation functions.

Perhaps the most important application of these ideas was to a proof of the conjectured amplitude/Wilson loop duality.  This conjecture, originally due to Alday and Maldacena,  arose from AdS/CFT considerations and stated that the planar amplitude (i.e., in the large $N$ limit for gauge group $SU(N)$) for maximally supersymmetric Yang Mills is equivalent to a Wilson loop around a null polygon whose sides are made from the momenta of the particles in the scattering amplitude; the planarity condition means that there is a trace order for the gluons in the amplitude as in (\ref{pt_amp}) that determines the ordering of the momenta around the polygon.  This null polygon is particularly easy to represent in twistor space, as it corresponds naturally to a generic polygon there and one can express the Wilson-loop on space-time in terms of a holomorphic Wilson-loop in twistor space that can be computed via the twistor action.  
In the Feynman diagram framework that arises from the twistor action in the axial gauge, one finds that the Feynman diagrams for the holomorphic  Wilson loop are dual to those for the amplitude in the sense of planar duality, giving a proof of the amplitude/Wilson loop correspondence at the level of the loop integrand \cite{Mason:2010yk}. 

Much of this and related material is reviewed in \cite{adamo_review} although this does not cover more recent work on stress-energy correlators \cite{Chicherin:2014uca} and form factors \cite{ Koster:2016loo, Chicherin:2016qsf}.

\subsection{Recursion relations, Grassmannians and Polytopes}

The BCFW recursion relations \cite{Britto:2005fq} were  a separate major development that sprang from the twistor string.  These use an elegant Cauchy theorem argument to construct tree amplitudes with $n$ external particles  from  amplitudes with fewer external particles. The idea is to introduce a complex parameter $z$ by shifting the spinor representation of the momenta of  two of the particles
$$
(\pi_1^{A'}\tilde\pi_1^A,\pi_n^{A'}\tilde\pi_n^A)\rightarrow (\pi_1^{A'}(\tilde\pi_1^A+z\tilde\pi_n^A),(\pi_n^{A'}-z\pi^{A'}_1)\tilde\pi_n^A)\, .
$$
As a function of $z$, an amplitude only has simple poles where the momenta flowing through an internal propagator becomes null (i.e., where some partial sum of momentum becomes null as $z$ varies).  The residues at these poles are products of the tree amplitudes on each side of the propagator evaluated at the shifted momentum.  Thus the amplitude can be expressed as the residue of $1/z$ times the amplitude at $z=0$, but this can be expressed as the sum of the residues  where $z\neq 0$ which are expressed in terms of lower order amplitudes (it turns out that there are many good situations where there is no contribution from $z=\infty$). These recursion formulae have subsequently been extended quite widely to include gravity and Yang-Mills in various dimensions and with varying amount of supersymmetry \cite{ArkaniHamed:2008gz}, and to loop integrands of planar gauge theories \cite{ArkaniHamed:2010kv}.
Andrew Hodges \cite{Hodges:2005bf} expressed the recursion relations in terms of twistor diagrams providing a generating principle that had hitherto been missing. 

Further work on expressing BCFW recursion in twistor space \cite{ArkaniHamed:2009si,Mason:2009sa} led to a Grassmannian contour  formula for amplitudes and leading singularities (invariants of multiloop amplitudes) \cite{ArkaniHamed:2009dn}.  A related Grassmannian formula was obtained soon after, which with hindsight, gave the  analagous Grassmannian contour integral formulae for the Wilson-loop, but, by the  amplitude/Wilson-loop duality, gives an alternative but quite different contour integral formula for amplitudes \cite{MS09} that are rather simpler than the original Grassmannian formulae.

This led to a programme to understand the residues that arise in the Grassmannian.  It emerged that a key idea that is required is to think of the Grassmannians as real manifolds and study their positive geometry \cite{nima1} leading to simple combinatorial characterisations of the residues.

The twistors for the Wilson-loop version of the amplitude were first introduced by Hodges  \cite{Hodges:2009hk} and called momentum twistors as they can be obtained locally from the momenta for the amplitude.  He used them in a completely novel way to show that the redundancy and choices built into a BCFW decomposition of an amplitude could be understood geometrically in momentum twistor space as arising from different dissections of a polyhedron that describes the whole amplitude as one global object.  This was originally only realized for NMHV amplitudes, but the programme has continued and matured into the \emph{amplituhedron} \cite{nima2} (although this is essentially the dual of Hodges' original picture which has still not been fully realized).

\section{Gravitization of Quantum Mechanics and Newtonian Twistors}
\label{sec_newton}
In \cite{Pe96,Pe11} Roger Penrose has argued that
a collapse of the wave function is a real process taking place in time, and is 
not described by the Schr\"odinger equation. Gravity should have a role to play
in explaining the nature of the quantum collapse, and the conventional views on quantum mechanics
may need to be revisited in the process. The key idea is that
the superposition of massive states must correspond to  superposition
of space--times. This makes the notion of the stationary states ambiguous - its definition depends
on a choice of a time--like Killing vector. In \cite{Pe96} an essentially Newtonian calculation led to the conclusion that the timescale of instability of one
stationary state is
$
\tau\approx \hbar/E_G
$
where
$E_G$ is the gravitational energy need to separate two mass distributions.
To attempt a twistor understanding of this relation one first has to take a Newtonian limit of 
the {\em relativistic} twistor correspondence. Analysing the incidence relation
 (\ref{4d_incidence})
in the flat case shows that such a limit corresponds to the jumping line phenomenon \cite{DG16}. If ${\mathbb{PT}}_c$ is a family
of twistor spaces corresponding  to a flat Minkowski space parametrised
by a finite speed of light $c$, then exploiting the holomorphic fibration
${\mathbb{PT}}_c\rightarrow {\OO}(2)$ one finds
\[
\lim_{c\rightarrow\infty} ({\mathbb{PT}}_c={\OO}(1)\oplus{\OO}(1))=
{\mathbb{PT}}_{\infty}={\OO}\oplus{\OO}(2).
\]
In the Nonlinear Graviton and Ward correspondences (Theorems
\ref{theo3} and \ref{main_twistor_theorem}) the presence of jumping lines
corresponds to singularities of the metric  or gauge fields
on a hypersurface \cite{tod_h, H2, sparlingg}. In the Newtonian twistor theory
all curves jump.

The curved twistor space in the Newtonian limit can be understood 
by considering a one--parameter family of Gibbons--Hawking metrics 
(\ref{GH_metric})
\[
g(c)=(1+c^{-2} V)(dx^2+dy^2+dz^2)+{c^{2}
(1+c^{-2} V)^{-1}}(d\tau+
c^{-3} A)^2.
\]
and taking a limit $c\rightarrow \infty$ in the Gibbons--Hawking twistor space.
This leads to a Newton--Cartan theory \cite{Cartan,Trautman} on the moduli space $M$
of ${\cal O}(2)\oplus{\cal O}$ curves.  
The limit of $\{  c^{-2} g(c),  g^{-1}(c), \nabla(c)\}$, 
where $\nabla(c)$ is the Levi--Civita connection of $g(c)$ is a triple consisting of one--form $\theta=d\tau$ giving a fibration $M\rightarrow \R$
(absolute time), a degenerate inverse metric $h^{ij}=\delta^{ij}$ on the fibres,
and a torsion--free connection preserving $h$ and $\theta$ with the only nonvanishing
Christoffel symbols given by ${\Gamma^{i}}_{\tau\tau}=
\frac{1}{2}\delta^{ij}{\p_j V}$.
\begin{center}
\includegraphics[width=7cm,height=3cm,angle=0]{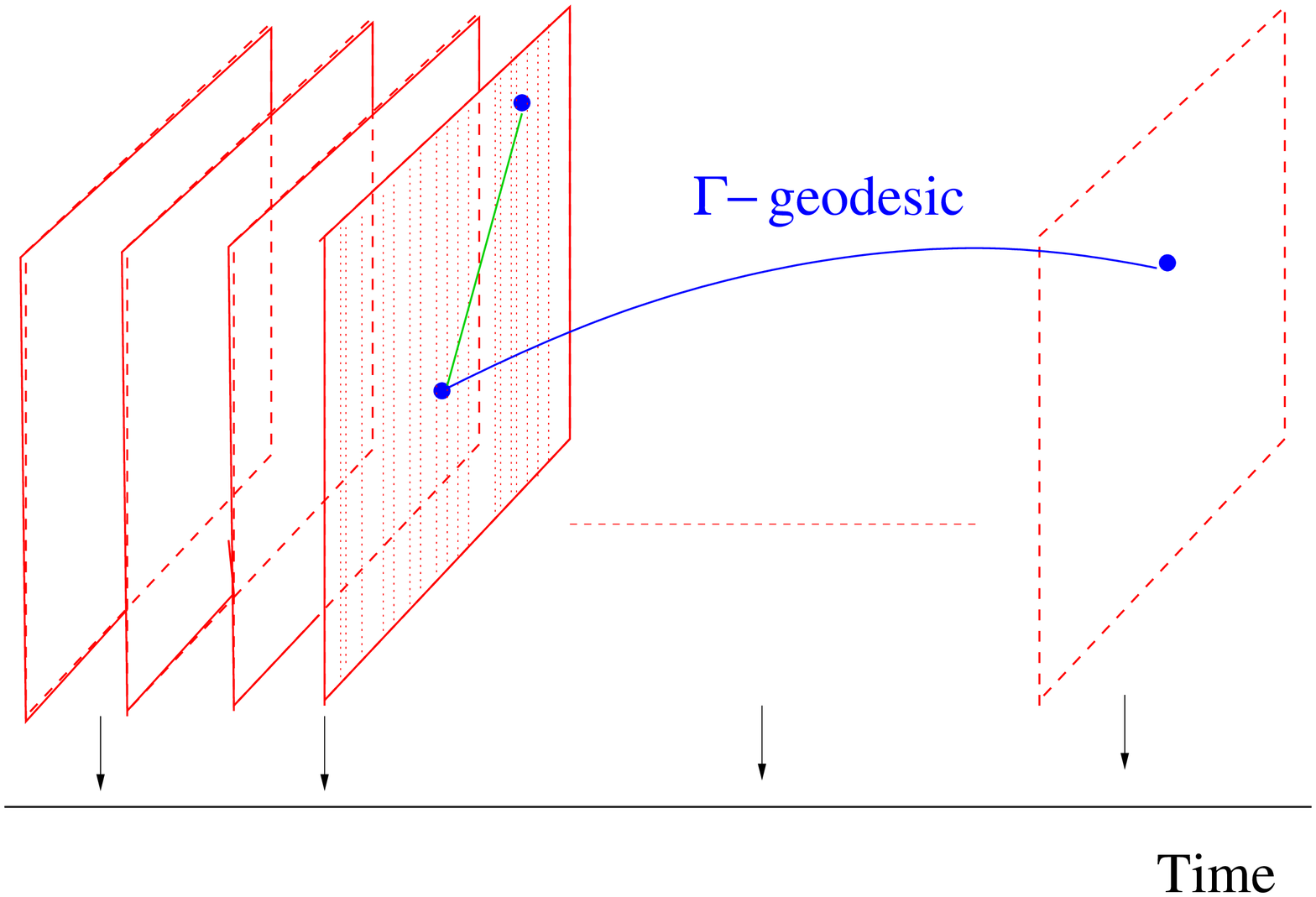}
\begin{center}
{{\bf Figure 6.} {\em Newton Cartan space-time.}}
\end{center}
\end{center}
\section{Other developments}
Our presentation has omitted many interesting applications of twistor methods.
A good reference are articles which appeared in Twistor Newsletter published
by the Twistor group in Mathematical Institute between 1976 and 2000,
and reprinted in \cite{advances,fadvances}. In particular they contain 
a discussion of the notoriously difficult and still unsolved {\em Googly Problem} of encoding self--dual (rather than anti--self--dual) and more general Einstein spaces in a geometry of ${\mathbb{PT}}$. See also
 \cite{twistor_in_maths_phys, law_thesis, devon_proc, penrose_light}
 (and \cite{Anew1, Anew2} for recent ideas of how four--manifolds can be used in constructing some geometric models of matter). 
Below we list some other developments.
\vskip3pt
 In {\em integral geometry} the role of space--time and twistor space is turned round.
The subject goes back to 
F. John \cite{john} who considered the problem
of characterising functions $\phi$ on the space or oriented lines
in $\R^3$ such that  $\phi(L):=\int_L f$ if $f:\R^3\rightarrow\R$. The image of this integral transform is characterised by the kernel of the ultra-hyperbolic Laplacian. The resulting integral
formula (see \cite{john, woodhouse_john}) is an analytic continuation
of Penrose's contour integral formula (\ref{contour_form}). 
A general relationship between  twistor theory and integral geometry has  
been explored in \cite{gindikin, novikov1, funk}
and in the language of systems of 2nd order ODEs in 
\cite{Gro, casey}.
\vskip3pt}
{\em Quasi local mass.} In the study of asymptotic properties of space--time one seeks
satisfactory definitions of energy and momentum which make sense at  extended, but finite 
regions of space--time, i. e. at the quasi local level. One definition associates these quantities
to an arbitrary two--surface in space time, and makes use of twistor theory of such surfaces
\cite{QM1, QM2, QM3}. Another twistor approach to this problem \cite{mf}
is based on the Ashtekar variables \cite{ash1}.
\vskip3pt
 In {\em loop quantum gravity} twistors provide a description of
spin network states. This approach has been developed by Simone Speziale and his collaborators.
See \cite{FS, LS}. Their description of symplectic structures and canonical quantisation
builds on a work of Tod \cite{tod_s}.
\vskip3pt
 Space-time points are derived objects in twistor theory. They become `fuzzy' after quantisation which initially seemed to be an attractive framework for quantum gravity. One possible realisation of this may be a {\em non--commutative twistor theory} \cite{Takasaki, przan, kapustin, hannabuss, majid, Lukier}
as well as the most recent twistorial contribution from Roger Penrose 
\cite{palatian}.

\section{Conclusions}
Twistor theory is a set of  non-local constructions with roots in 19th century  projective geometry. By now twistor ideas have been extended and generalized in many different directions and applied to many quite different problems in mathematics and physics.  A unifying feature
is the correspondence between points in space-time and holomorphic
curves or some family of higher dimensional compact complex submanifolds in a twistor space, together with the encoding of space-time data into some deformed complex structure. 

 Complex numbers play an essential role. The local non--linearities of 
anti--self--dual Einstein and 
Yang--Mills equations in space--time are replaced by algebro--geometric problems
in twistor--space or by the Cauchy--Riemann equations in the Atiyah--Hitchin--Singer picture
adapted to Riemannian reality conditions. 
Twistor and ambitwistor string theories couple directly to the complex geometry of twistor and ambitwistor spaces; the cohomology classes on twistor and ambitwistor spaces that represent space-time fields restrict directly to give the vertex operators of the corresponding theories.  These give a coherent twistorial formulation of many of the physical theories 
that lead to striking simplifications over their space-time formulations, particularly in amplitude calculations.
There is now clear evidence that the conventional string is tied up with the geometry of twistors in ten dimensions \cite{Berkovits:2014aia}.

We wish twistor theory and its founder
all the best on the occasion of their respective anniversaries.
We expect the future of twistor theory to be at least as productive as its past, and that there will much to celebrate in this 21st century.


\vskip2pc






\end{document}